\documentclass[11pt]{article}
\usepackage{graphicx}
\usepackage{tabularx}
\usepackage{url}
\usepackage{amsthm}
\usepackage{amsmath}
\usepackage{amsfonts}
\usepackage[labelfont=bf]{caption}

\theoremstyle{plain}
\newtheorem{theorem}{Theorem}
\newtheorem{lemma}{Lemma}

\theoremstyle{definition}
\newtheorem{definition}{Definition}

\theoremstyle{remark}
\newtheorem{remark}{Remark}

\newtheorem{exmp}{Example}

\date{}

\begin{document}

\title{Bounds on Depth of Decision Trees Derived from Decision Rule Systems}

\author{Kerven Durdymyradov and Mikhail Moshkov \\
Computer, Electrical and Mathematical Sciences \& Engineering Division \\ and Computational Bioscience Research Center\\
King Abdullah University of Science and Technology (KAUST) \\
Thuwal 23955-6900, Saudi Arabia\\ \{kerven.durdymyradov,mikhail.moshkov\}@kaust.edu.sa
}

\maketitle

\begin{abstract}
Systems of decision rules and decision trees are widely used as a means for knowledge representation, as classifiers, and as algorithms. They are among the most interpretable models for classifying and representing knowledge. The study of relationships between these two models is an important task of computer science. It is easy to transform a decision tree into a decision rule system. The inverse transformation is a more difficult task. In this paper, we study unimprovable upper and lower bounds on the minimum depth of decision trees derived from decision rule systems depending on the various parameters of these systems.
\end{abstract}

\section{Introduction\label{S1}}

Decision trees \cite%
{AbouEishaACHM19,AlsolamiACM20,BreimanFOS84,Moshkov05,MoshkovZ11,RokachM07}
and decision rule systems \cite%
{BorosHIK97,BorosHIKMM00,ChikalovLLMNSZ13,FurnkranzGL12,MPZ08,MoshkovZ11,Pawlak91,PawlakS07}
are widely used as a means for knowledge representation, as classifiers that
predict decisions for new objects, and as algorithms for solving various
problems of fault diagnosis, combinatorial optimization, etc. Decision trees
and rules are among the most interpretable models for classifying and
representing knowledge \cite{CaoSJ20,GilmoreEH21,Molnar22,SilvaGKJS20}.

The study of relationships between decision trees and systems of decision
rules is an important task of computer science. Regular methods for
extracting decision rules from decision trees have been known for a long
time \cite{Quinlan87,Quinlan93,Quinlan99}: first, each path in the decision
tree from the root to a terminal node is assigned a decision rule, and then
the resulting rules are simplified. In this paper, we study the more complex
problem of transforming a decision rule system into a decision tree. There
are at least three directions of research related to this problem:

\begin{itemize}
\item In a number of papers \cite%
{AbdelhalimTN16,AbdelhalimTS09,ImamM93a,ImamM93,ImamM96,KaufmanMPW06,MichalskiI94,MichalskiI97,SzydloSM05}%
, it was proposed to build decision trees or their generalizations, known as
decision structures, in two stages: first, build decision rules based on the
input data, and then build decision trees or decision structures based on
the constructed rules. There are various methodological considerations and
experimental results that justify this approach.

\item Relations between the depth of deterministic and nondeterministic
decision trees were studied both for decision tables and for problems on
finite and infinite information systems, each of which consists of a
universe and a set of attributes defined on it \cite%
{Moshkov96,Moshkov00,Moshkov03,Moshkov05a,Moshkov20}. A nondeterministic
decision tree can be interpreted as a system of true decision rules for the
given table (problem) that covers all rows (inputs). The most famous results
in this direction are related to decision trees for computing Boolean
functions \cite{BlumI87,HartmanisH87,Moshkov95,Tardos89}. In this case, the
minimum depth of a nondeterministic decision tree for a Boolean function $f$
is equal to the certificate complexity of $f$ \cite{BuhrmanW02}.

\item One of the authors of this paper begun in \cite{Moshkov98,Moshkov01}
the development of the so-called syntactic approach to the study of the
considered problem, which assumes that we do not know input data but only
have a system of decision rules that must be transformed into a decision tree.
This paper continues work in this direction. The results obtained in \cite%
{Moshkov98,Moshkov01} are described in remarks after Theorems 1 and 2.
\end{itemize}

Let there be a system of decision rules $S$ of the form $(a_{i_{1}}=\delta
_{1})\wedge \cdots \wedge (a_{i_{m}}=\delta _{m})\rightarrow \sigma $, where
$a_{i_{1}},\ldots ,a_{i_{m}}$ are attributes, $\delta _{1},\ldots ,\delta
_{m}$ are values of these attributes, and $\sigma $ is a decision. We
describe three problems associated with this system:

\begin{itemize}
\item For a given input (a tuple of values of all attributes included in $S$%
), it is necessary to find at least one rule that is realizable for this
input (having a true left-hand side) or show that there are no such rules.

\item For a given input, it is necessary to find all the rules that are
realizable for this input, or show that there are no such rules.

\item For a given input, it is necessary to find all the right-hand sides of
rules that are realizable for this input, or show that there are no such
rules.
\end{itemize}

For each problem, we consider two its variants. The first assumes that in
the input each attribute can have only those values that occur for this
attribute in the system $S$. In the second case, we assume that in the input
any attribute can have any value.

Our goal is to minimize the number of queries for attribute values in the
given input. For this purpose, decision trees are studied as algorithms for
solving the considered six problems.

In this paper, we investigate for each of these problems, unimprovable upper and lower bounds on the minimum depth of decision trees depending on three parameters of the decision rule system – the total number of different attributes in the rules belonging to the system, the maximum length of the decision rule, and the maximum number of attribute values.

We show that, for each problem, there are systems of decision rules for which the minimum depth of the decision trees that solve the problem is much less than the total number of attributes in the rule system. For such systems of decision rules, it is advisable to use decision trees.

This paper consists of six sections. Section \ref{S2} discusses the main
definitions and notation. Sections \ref{S3}-\ref{S5} are devoted to the study of unimprovable lower and upper bounds on the depth of decision trees depending on the parameters
of the system of decision rules: Section \ref{S3} contains auxiliary statements and Sections \ref{S4} and \ref{S5} contain proofs of upper and lower bounds, respectively. Section \ref{S6} contains a short conclusion.

\section{Main Definitions and Notation\label{S2}}

In this section, we discuss the main definitions and notation related to
decision rule systems, decision trees, and functions that characterize the depth of decision trees derived from decision rule systems.

\subsection{Decision Rule Systems\label{S2.1}}

Let $\omega =\{0,1,2,\ldots \}$ and $A=\{a_{i}:i\in \omega \}$. Elements of
the set $A$ will be called \emph{attributes}. 

\begin{definition}
A \emph{decision rule} is an expression of the form
\begin{equation*}
(a_{i_{1}}=\delta _{1})\wedge \cdots \wedge (a_{i_{m}}=\delta
_{m})\rightarrow \sigma ,
\end{equation*}%
where $m\in \omega $, $a_{i_{1}},\ldots ,a_{i_{m}}$ are pairwise different
attributes from $A$ and $\delta _{1},\ldots ,\delta _{m},\sigma \in \omega $. 
\end{definition}

We denote this decision rule by $r$. The expression $(a_{i_{1}}=\delta
_{1})\wedge \cdots \wedge (a_{i_{m}}=\delta _{m})$ will be called the \emph{%
left-hand side}, and the number $\sigma $ will be called the \emph{%
right-hand side} of the rule $r$. The number $m$ will be called the \emph{%
length }of the decision rule $r$. Denote $A(r)=\{a_{i_{1}},\ldots
,a_{i_{m}}\}$ and $K(r)=\{a_{i_{1}}=\delta _{1},\ldots ,a_{i_{m}}=\delta
_{m}\}$. If $m=0$, then $A(r)=K(r)=\emptyset $.

\begin{definition}
Two decision rules $r_1$ and $r_2$ are \emph{equal} if $K(r_1)=K(r_2)$ and the right-hand sides of the rules $r_1$ and $r_2$ are equal.
\end{definition}

\begin{definition}
A \emph{system of decision rules} $S$ is a finite nonempty set of decision
rules. 
\end{definition}

Denote $A(S)=\bigcup_{r\in S}A(r)$, $n(S)=\left\vert A(S)\right\vert $%
, $D(S)$ the set of the right-hand sides of decision rules from $S$, and $%
d(S)$ the maximum length of a decision rule from $S$. Let $n(S)>0$. For $a_{i}\in A(S)$,
let $V_{S}(a_{i})=\{\delta :a_{i}=\delta \in \bigcup_{r\in S}K(r)\}$ and $%
EV_{S}(a_{i})=V_{S}(a_{i})\cup \{\ast \}$, where the symbol $\ast $ is
interpreted as a number that does not belong to the set $V_{S}(a_{i})$.
Denote $k(S)=\max \{\left\vert V_{S}(a_{i})\right\vert :a_{i}\in A(S)\}$. If $n(S)=0$, then $k(S)=0$. We
denote by $\Sigma $ the set of systems of decision rules.

\begin{exmp}
Let us consider a decision rule system $S=\{(a_{1}=0)\wedge (a_{2}=0)\wedge (a_{3}=0)\rightarrow 3, (a_{1}=1)\wedge (a_{4}=0)\rightarrow 4, (a_{1}=2)\rightarrow 5\}$. Then $A(r_1)=\{a_{1},a_{2},a_{3}\}$, $K(r_1)=\{a_{1}=0,a_{2}=0,a_{3}=0\}$, where $r_1$ denotes the first rule from $S$. $A(S)=\bigcup_{r\in S}A(r) = \{a_{1},a_{2},a_{3},a_{4}\}$, $n(S)=\left\vert A(S)\right\vert = 4$, $D(S) = \{3,4,5\}$, $d(S) = 3$, $V_{S}(a_{1})=\{\delta :a_{1}=\delta \in \bigcup_{r\in S}K(r)\} = \{0, 1, 2\}$, $EV_{S}(a_{1})=V_{S}(a_{1})\cup \{\ast \} = \{0, 1, 2, \ast\}$ and $k(S)=\max \{\left\vert V_{S}(a_{i})\right\vert :a_{i}\in A(S)\} = \left\vert V_{S}(a_{1})\right\vert = 3$.
\end{exmp}

Let $S\in \Sigma $, $n(S)>0$, and $A(S)=\{a_{j_{1}},\ldots ,a_{j_{n}}\}$, where $%
j_{1}<\cdots <j_{n}$. Denote $V(S)=V_{S}(a_{j_{1}})\times \cdots \times
V_{S}(a_{j_{n}})$ and $EV(S)=EV_{S}(a_{j_{1}})\times \cdots \times
EV_{S}(a_{j_{n}})$. For $\bar{\delta}=(\delta _{1},\ldots ,\delta _{n})\in
EV(S)$, denote $K(S,\bar{\delta})=\{a_{j_{1}}=\delta _{1},\ldots
,a_{j_{n}}=\delta _{n}\}$. 

\begin{definition}
We will say that a decision rule $r$ from $S$ is \emph{%
realizable} for a tuple $\bar{\delta} \in EV(S)$ if $K(r)\subseteq K(S,\bar{\delta})$. 
\end{definition}

It is clear that any rule with an empty left-hand side is realizable for the tuple $\bar{\delta}$.

\begin{exmp}
Let us consider a decision rule system $S  = \{r_1: (a_{1}=0)\wedge (a_{2}=0)\rightarrow 0$, $r_2: (a_{1}=0)\wedge (a_{3}=1)\rightarrow 1\}$ and a tuple $\bar{\delta} = (0,0,0) \in EV(S)$. Then the decision rule $r_1$ from $S$ is realizable for the tuple $\bar{\delta}$, but $r_2$ is not.
\end{exmp}

Let $V\in \{V(S),EV(S)\}$. We now define three problems related to the rule
system $S$.

\begin{definition}
Problem \emph{All Rules} for the pair $(S,V)$: for a given tuple $\bar{\delta%
}\in V$, it is required to find the set of rules from $S$ that are
realizable for the tuple $\bar{\delta}$.
\end{definition}

\begin{definition}
Problem \emph{All Decisions} for the pair $(S,V)$: for a given tuple $\bar{%
\delta}\in V$, it is required to find a set $Z$ of decision rules from $S$
satisfying the following conditions:

\begin{itemize}
\item All decision rules from $Z$ are realizable for the tuple $\bar{\delta}$%
.

\item For any $\sigma \in D(S)\setminus D(Z)$, any decision rule from $S$
with the right-hand side equal to $\sigma $ is not realizable for the tuple $%
\bar{\delta}$.
\end{itemize}
\end{definition}

\begin{definition}
Problem \emph{Some Rules} for the pair $(S,V)$: for a given tuple $\bar{%
\delta}\in V$, it is required to find a set $Z$ of decision rules from $S$
satisfying the following conditions:

\begin{itemize}
\item All decision rules from $Z$ are realizable for the tuple $\bar{\delta}$%
.

\item If $Z=\emptyset $, then any decision rule from $S$ is not realizable
for the tuple $\bar{\delta}$.
\end{itemize}
\end{definition}

Denote $AR(S)$ and $EAR(S)$ the problems All Rules for pairs $(S,V(S))$ and $%
(S,EV(S))$, respectively. Denote $AD(S)$ and $EAD(S)$ the problems All
Decisions for pairs $(S,V(S))$ and $(S,EV(S))$, respectively. Denote $SR(S)$
and $ESR(S)$ the problems Some Rules for pairs $(S,V(S))$ and $(S,EV(S))$,
respectively.

\begin{exmp}
Let a decision rule system $S=\{(a_{1}=0)\wedge (a_{2}=0)\wedge (a_{3}=0)\rightarrow 0, (a_{1}=0)\wedge (a_{2}=0)\rightarrow 1, (a_{1}=0)\rightarrow 0\}$ and a tuple $\bar{\delta}=(0,0,0) \in EV(S)$ are given. Then $\{(a_{1}=0)\wedge (a_{2}=0)\wedge (a_{3}=0)\rightarrow 0, (a_{1}=0)\wedge (a_{2}=0)\rightarrow 1, (a_{1}=0)\rightarrow 0\}$ is the solution for the problem $AR(S)$ and the tuple $\bar{\delta}$, $\{(a_{1}=0)\wedge (a_{2}=0)\rightarrow 1, (a_{1}=0)\rightarrow 0\}$ is a solution for the problem $AD(S)$ and the tuple $\bar{\delta}$, and $\{(a_{1}=0)\rightarrow 0\}$ is a solution for the problem $SR(S)$ and the tuple $\bar{\delta}$.
\end{exmp}

In the special case, when $n(S)=0$,  all rules from $S$ have an empty left-hand side. In this case, it is natural to consider (i) the set $S$ as the solution to the problems $AR(S)$ and $EAR(S)$, (ii) any subset $Z$ of the set $S$ with $D(Z)=D(S)$ as a solution to the problems $AD(S)$ and $EAD(S)$, and (iii) any nonempty subset $Z$ of the set $S$ as a solution to the problems $SR(S)$ and $ESR(S)$.

\subsection{Decision Trees\label{S2.2}}

A \emph{finite directed tree with root} is a finite directed tree in which
only one node has no entering edges. This node is called the \emph{root}.
The nodes without leaving edges are called \emph{terminal} nodes. The nodes
that are not terminal will be called \emph{working} nodes. A \emph{complete
path} in a finite directed tree with root is a sequence $\xi
=v_{1},d_{1},\ldots ,v_{m},d_{m},v_{m+1}$ of nodes and edges of this tree in
which $v_{1}$ is the root, $v_{m+1}$ is a terminal node and, for $i=1,\ldots
,m$, the edge $d_{i}$ leaves the node $v_{i}$ and enters the node $v_{i+1}$.

We will consider two types of decision trees: o-decision trees (ordinary
decision trees, o-trees in short) and e-decision trees (extended decision
trees, e-trees in short). 

\begin{definition}
A \emph{decision tree over a decision rule system}
$S$ is a labeled finite directed tree with root $\Gamma $ satisfying the
following conditions:

\begin{itemize}
\item Each working node of the tree $\Gamma $ is labeled with an attribute
from the set $A(S)$.

\item Let a working node $v$ of the tree $\Gamma $ be labeled with an
attribute $a_{i}$. If $\Gamma $ is an o-tree, then exactly $\left\vert
V_{S}(a_{i})\right\vert $ edges leave the node $v$ and these edges are
labeled with pairwise different elements from the set $V_{S}(a_{i})$. If $%
\Gamma $ is an e-tree, then exactly $\left\vert EV_{S}(a_{i})\right\vert $
edges leave the node $v$ and these edges are labeled with pairwise different
elements from the set $EV_{S}(a_{i})$.

\item Each terminal node of the tree $\Gamma$ is labeled with a subset of
the set $S$.
\end{itemize}
\end{definition}

Let $\Gamma $ be a decision tree over the decision rule system $S$. We
denote by $CP(\Gamma )$ the set of complete paths in the tree $\Gamma $. Let
$\xi =v_{1},d_{1},\ldots ,v_{m},d_{m},v_{m+1}$ be a complete path in $\Gamma
$. We correspond to this path a set of attributes $A(\xi )$ and an equation
system $K(\xi )$. If $m=0$ and $\xi =v_{1}$, then $A(\xi )=\emptyset $ and $K(\xi )=\emptyset
$. Let $m>0$ and, for $j=1,\ldots ,m$, the node $v_{j}$ be labeled with the
attribute $a_{i_{j}}$ and the edge $d_{j}$ be labeled with the element $%
\delta _{j}\in \omega \cup \{\ast \}$. Then $A(\xi )=\{a_{i_{1}},\ldots
,a_{i_{m}}\}$ and $K(\xi )=\{a_{i_{1}}=\delta _{1},\ldots ,a_{i_{m}}=\delta
_{m}\}$. We denote by $\tau (\xi )$ the set of decision rules attached to
the node $v_{m+1}$.

\begin{exmp}
Let us consider a decision rule system $S=\{(a_{1}=0)\wedge (a_{2}=0)\rightarrow 0, (a_{1}=1)\wedge (a_{3}=0)\rightarrow 0, (a_{1}=1)\rightarrow 0\}$. Then o-tree $\Gamma_1$ and e-tree $\Gamma_2$ over the decision rule system $S$ are given in Fig. \ref{fig:fig1}, where $r_1$, $r_2$ and $r_3$ are the first, second and third decision rules in $S$, respectively.

\begin{figure}[h!]
\renewcommand{\figurename}{\textbf{Fig.}}
    \centering
    \includegraphics[width=0.55\textwidth]{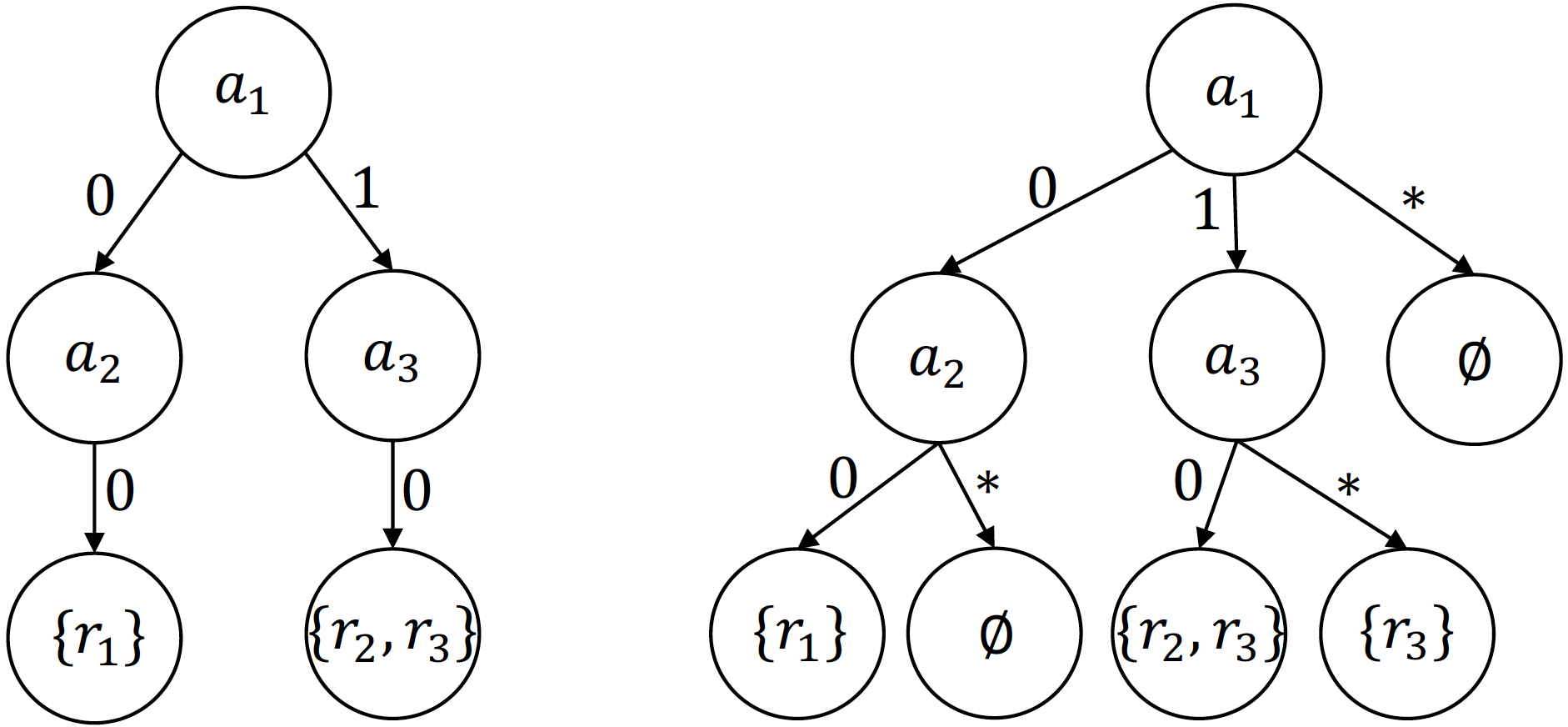}
    \caption{o-decision tree $\Gamma_1$ and e-decision $\Gamma_2$ tree over the decision rule system $S$}
    \label{fig:fig1}
\end{figure}
Let $\xi$ be a complete path in the o-tree, which is finished in the terminal node labeled with the set of rules $\{r_2, r_3\}$. Then $A(\xi )=\{a_{1}, a_{3}\}$,  $K(\xi )=\{a_{1}=1, a_{3}=0\}$ and  $\tau (\xi ) = \{r_{2}, r_{3}\}$.
\end{exmp}

\begin{definition}
A system of equations $\{a_{i_{1}}=\delta _{1},\ldots ,a_{i_{m}}=\delta
_{m}\}$, where $a_{i_{1}},\ldots ,a_{i_{m}}\in A$ and $\delta _{1},\ldots
,\delta _{m}\in \omega \cup \{\ast \}$, will be called \emph{inconsistent}
if there exist $l,k\in \{1,\ldots ,m\}$ such that $l\neq k$, $i_{l}=i_{k}$,
and $\delta _{l}\neq \delta _{k}$. If the system of equations is not
inconsistent then it will be called \emph{consistent}.
\end{definition}

Let $S$ be a decision rule system and $\Gamma $ be a decision tree over $S$.

\begin{definition}
We will say that $\Gamma $ \emph{solves} the problem $AR(S)$ (the problem $EAR(S)$, respectively) if $\Gamma $ is an o-tree (an e-tree, respectively)
and any path $\xi \in CP(\Gamma )$ with consistent system of equations $%
K(\xi )$ satisfies the following conditions:

\begin{itemize}
\item For any decision rule $r\in \tau (\xi )$, the relation $K(r)\subseteq
K(\xi )$ holds.

\item For any decision rule $r\in S\setminus \tau (\xi )$, the system of
equations $K(r)\cup K(\xi )$ is inconsistent.
\end{itemize}
\end{definition}

\begin{exmp}
Let $S$ be a decision rule system from Example 4. Then the decision trees $\Gamma_1$ and $\Gamma_2$ depicted in Fig. \ref{fig:fig1} solve the problems $AR(S)$ and $EAR(S)$, respectively.
\end{exmp}

\begin{definition}
We will say that $\Gamma $ \emph{solves} the problem $AD(S)$ (the problem $%
EAD(S)$, respectively) if $\Gamma $ is an o-tree (an e-tree, respectively)
and any path $\xi \in CP(\Gamma )$ with consistent system of equations $%
K(\xi )$ satisfies the following conditions:

\begin{itemize}
\item For any decision rule $r\in \tau (\xi )$, the relation $K(r)\subseteq
K(\xi )$ holds.

\item If $r\in S\setminus \tau (\xi )$ and the right-hand side of $r$ does
not belong to the set $D(\tau (\xi ))$, then the system of equations $%
K(r)\cup K(\xi )$ is inconsistent.
\end{itemize}
\end{definition}

\begin{definition}
We will say that $\Gamma $ \emph{solves} the problem $SR(S)$ (the problem $%
ESR(S)$, respectively) if $\Gamma $ is an o-tree (an e-tree, respectively)
and any path $\xi \in CP(\Gamma )$ with consistent system of equations $%
K(\xi )$ satisfies the following conditions:

\begin{itemize}
\item For any decision rule $r\in \tau (\xi )$, the relation $K(r)\subseteq
K(\xi )$ holds.

\item If $\tau (\xi )=\emptyset $, then, for any decision rule $r\in S$, the
system of equations $K(r)\cup K(\xi )$ is inconsistent.
\end{itemize}
\end{definition}

\begin{exmp}
Let us consider a decision rule system $S=\{(a_{1}=0)\wedge (a_{2}=0)\rightarrow 0, (a_{1}=1)\wedge (a_{3}=0)\rightarrow 1, (a_{1}=1)\rightarrow 1, (a_{1}=1)\rightarrow 2\}$ and decision trees $\Gamma_1$, $\Gamma_2$, and $\Gamma_3$ depicted in Fig. \ref{fig:fig2}. Then the decision tree $\Gamma_1$ solves the problem $AR(S)$, $\Gamma_2$ solves $AD(S)$, and $\Gamma_3$ solves $SR(S)$, where $r_1$, $r_2$, $r_3$ and $r_4$ are the first, second, third and fourth decision rules in $S$, respectively.

\begin{figure}[h!]
\renewcommand{\figurename}{\textbf{Fig.}}
    \centering
    \includegraphics[width=0.6\textwidth]{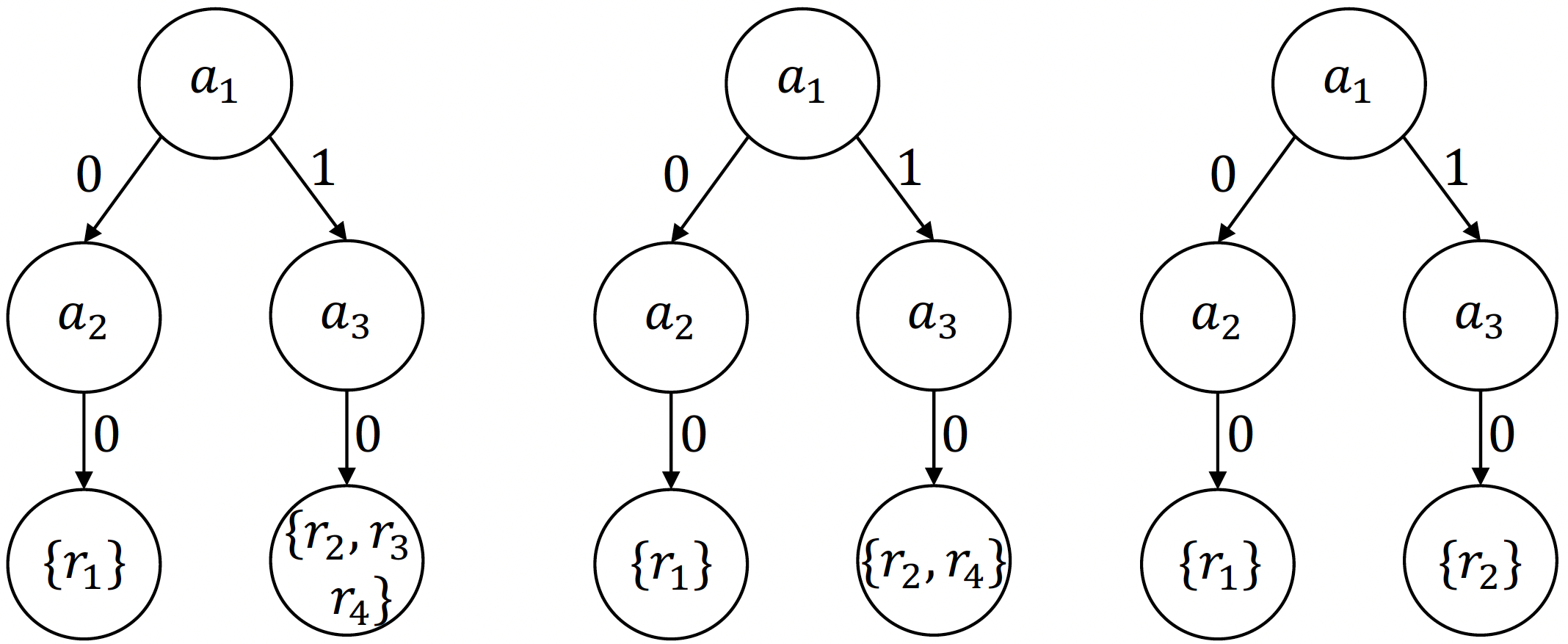}
    \caption{Decision trees $\Gamma_1$, $\Gamma_2$, and $\Gamma_3$}
    \label{fig:fig2}
\end{figure}

\end{exmp}

For any complete path $\xi \in CP(\Gamma )$, we denote by $h(\xi )$ the
number of working nodes in $\xi $. The value $h(\Gamma )=\max \{h(\xi ):\xi
\in CP(\Gamma )\}$ is called the \emph{depth} of the decision tree $\Gamma $.

Let $S$ be a decision rule system and $C\in \{AR,EAR,AD,EAD,SR,ESR\}$. We
denote by $h_{C}(S)$ the minimum depth of a decision tree over $S$,
which solves the problem $C(S)$.

Let $n(S)=0$. If $C \in \{AR,EAR\}$, then there is only one decision tree solving the problem $C(S)$. This tree consists of one node labeled with the set of rules $S$. If $C \in \{AD,EAD\}$, then the set of decision trees solving the problem $C(S)$ coincides with the set of trees each of which consists of one node labeled with a subset $Z$ of the set $S$ with $D(Z)=D(S)$. If $C \in \{SR,ESR\}$, then the set of decision trees solving the problem $C(S)$ coincides with the set of trees each of which consists of one node labeled with a nonempty subset $Z$ of the set $S$. Therefore if $n(S)=0$, then $h_{C}(S)=0$ for any $C\in \{AR,EAR,AD,EAD,SR,ESR\}$.

\subsection{Functions Characterizing Depth of Decision Trees\label{S2.3}}

Let $S\in \Sigma $, where $\Sigma $ is the set of decision rule systems. We
denote by $R_{SR}(S)$ a subsystem of the system $S$ that consists of all
rules $r\in S$ satisfying the following condition: there is no a rule $%
r^{\prime }\in S$ such that $K(r^{\prime })\subset K(r)$. 

\begin{definition}
The system $S$ will be called $SR$-\emph{reduced} if $R_{SR}(S)=S$.
\end{definition}

Denote by $\Sigma _{SR}$ the set of $SR$-reduced systems of decision rules.

For $S\in \Sigma $, we denote by $R_{AD}(S)$ a subsystem of the system $S$ that
consists of all rules $r\in S$ satisfying the following condition: there is
no a rule $r^{\prime }\in S$ such that $K(r^{\prime })\subset K(r)$ and the
right-hand sides of the rules $r$ and $r^{\prime }$ coincide. 

\begin{definition}
The system $S$ will be called $AD$-\emph{reduced} if $R_{AD}(S)=S$.
\end{definition}

Denote by $\Sigma _{AD}$ the set of $AD$-reduced systems of decision rules.

\begin{exmp}
Let us consider a decision rule system $S=\{(a_{1}=0)\wedge (a_{2}=0)\wedge (a_{3}=0)\rightarrow 0, (a_{1}=0)\wedge (a_{2}=0)\rightarrow 0, (a_{1}=0)\rightarrow 1\}$. For this system, $R_{AD}(S)=\{(a_{1}=0)\wedge (a_{0}=0)\rightarrow 0, (a_{1}=0)\rightarrow 1\}$ and $R_{SR}(S)=\{(a_{1}=0)\rightarrow 1\}$.
\end{exmp}

It is easy to show that%
\begin{eqnarray*}
\{(n(S),d(S),k(S)) :S\in \Sigma \}&=&\{(n(S),d(S),k(S)):S\in \Sigma _{SR}\}
\\
&=&\{(n(S),d(S),k(S)):S\in \Sigma _{AD}\} \\
&=&\{(0,0,0)\}\cup \{(n,d,k):n,d,k\in \omega \setminus \{0\},d\leq n\}.
\end{eqnarray*}%

Really, if $n(S)=0$, i.e., each rule from $S$ has empty left-hand side, then $d(S)=k(S)=0$. If $n(S)>0$, then $d(S)>0$, $k(S)>0$, and $d(S) \le n(S)$ since, in any decision rule, attributes from different equations in the left-hand side are different. Let $n,d,k\in \omega \setminus \{0\}$, $d\leq n$, and
$S=\{(a_1=0)\wedge \cdots \wedge (a_d=0)\rightarrow 0, (a_{d+1}=0)\rightarrow 0, \ldots , (a_n=0)\rightarrow 0, (a_1=1)\rightarrow 0, \ldots , (a_1=k-1)\rightarrow 0\}$.
One can show, that $S$ is $AD$-reduced and $SR$-reduced, $n(S)=n$,  $d(S)=d$, and $k(S)=k$.

We will not study the case, when $(n(S),d(S),k(S))=(0,0,0)$, since if $n(S)=0$, then $h_{C}(S)=0$ for any $C\in \{SR,ESR,AD,EAD,AR,EAR\}$.

\begin{definition}
Let $C\in \{SR,ESR,AD,EAD,AR,EAR\}$, $n,d,k\in \omega \setminus \{0\}$, and $%
d\leq n$. Denote%
\begin{eqnarray*}
h_{C}(n,d,k) &=&\min \{h_{C}(S):S\in \Sigma ,n(S)=n,d(S)=d,k(S)=k\}, \\
H_{C}(n,d,k) &=&\max \{h_{C}(S):S\in \Sigma ,n(S)=n,d(S)=d,k(S)=k\}.
\end{eqnarray*}
\end{definition}

The considered parameters are lower ($h_{C}(n,d,k)$) and upper ($%
H_{C}(n,d,k) $) unimprovable bounds on the minimum depth of decision trees
solving the problem $C(S)$ for systems of decision rules $S$ such that $%
n(S)=n $, $d(S)=d$, and $k(S)=k$.

We also consider similar parameters for $SR$-reduced and for $AD$%
-reduced systems of decision rules. Interest in the study of such rule
systems is due to the fact that the transformation of a system $S$ into
systems $R_{SR}(S)$ and $R_{AD}(S)$ is not difficult and the systems $%
R_{SR}(S)$ and $R_{AD}(S)$ can be much simpler than the system $S$. In
addition, as Lemma \ref{L9} shows, equalities $h_{ESR}(S)=
h_{ESR}(R_{SR}(S))$ and $h_{EAD}(S)=h_{EAD}(R_{AD}(S))$ hold, and the
transformation of decision trees solving problems $ESR(R_{SR}(S))$ and $%
EAD(R_{AD}(S))$ into decision trees solving problems $ESR(S)$ and $EAD(S)$,
respectively, can be carried out relatively simply.

\begin{definition}
Let $C\in \{SR,ESR,AD,EAD\}$, $n,d,k\in \omega \setminus \{0\}$, and $d\leq
n $. Let $C^{\prime }=SR$ if $C\in \{SR,ESR\}$ and $C^{\prime }=AD$ if $C\in
\{AD,EAD\}$. Denote%
\begin{eqnarray*}
h_{C}^{R}(n,d,k) &=&\min \{h_{C}(S):S\in \Sigma _{C^{\prime
}},n(S)=n,d(S)=d,k(S)=k\}, \\
H_{C}^{R}(n,d,k) &=&\max \{h_{C}(S):S\in \Sigma _{C^{\prime
}},n(S)=n,d(S)=d,k(S)=k\}.
\end{eqnarray*}
\end{definition}

The considered parameters are lower $(h_{C}^{R}(n,d,k))$ and upper $(H_{C}^{R}(n,d,k))$ unimprovable bounds on the minimum depth of decision trees solving the problem $C(S)$ for $C^{\prime }$-reduced systems of decision rules $S$ such that $n(S) = n$, $d(S) = d$, and $k(S) = k$.

\section{Auxiliary Statements\label{S3}}

In this section, we prove several auxiliary statements that will be used later.

Let $S$ be a decision rule system and $\Gamma $ be an e-decision tree over $%
S $. We denote by $o(\Gamma )$ an o-tree over $S$, which is obtained from
the tree $\Gamma $ by the removal of all nodes $v$ such that the path from
the root to the node $v$ in $\Gamma $ contains an edge labeled with $\ast $.
Together with a node $v$, we remove all edges entering or leaving $v$. One
can prove the following two lemmas.

\begin{lemma}
\label{L1} Let $S$ be a decision rule system, $C\in \{AR,AD,SR\}$, and $%
\Gamma $ be an e-decision tree over $S$ solving the problem $EC(S)$. Then the
decision tree $o(\Gamma )$ solves the problem $C(S)$.
\end{lemma}

\begin{lemma}
\label{L2} Let $S$ be a decision rule system and $\Gamma $ be a decision
tree over $S$. Then

(a) If the tree $\Gamma $ solves the problem $AR(S)$ ($EAR(S)$,
respectively), then the tree $\Gamma $ solves the problem $AD(S)$ ($EAD(S)$,
respectively).

(b) If the tree $\Gamma $ solves the problem $AD(S)$ ($EAD(S)$,
respectively), then the tree $\Gamma $ solves the problem $SR(S)$ ($ESR(S)$,
respectively).
\end{lemma}

\begin{lemma}
\label{L3} Let $S$ be a decision rule system. Then the following
inequalities hold:

\begin{equation*}
\begin{array}{ccccccc}
h_{ESR}(S) & \leq & h_{EAD}(S) & \leq & h_{EAR}(S) & \leq & n(S) \\
\mathrel{\rotatebox{90}{$\le$}} &  & \mathrel{\rotatebox{90}{$\le$}} &  & %
\mathrel{\rotatebox{90}{$\le$}} &  &  \\
h_{SR}(S) & \leq & h_{AD}(S) & \leq & h_{AR}(S) &  &
\end{array}%
\end{equation*}
\end{lemma}

\begin{proof}
It is clear that the considered inequalities hold if $n(S)=0$. Let $n(S)>0$.
Let $\Gamma $ be an e-decision tree over $S$. It is clear that $h(o(\Gamma
))\leq h(\Gamma )$. Using this inequality and Lemma \ref{L1} we obtain $%
h_{SR}(S)\leq h_{ESR}(S)$, $h_{AD}(S)\leq h_{EAD}(S)$, and $h_{AR}(S)\leq
h_{EAR}(S)$. By Lemma \ref{L2}, $h_{ESR}(S)\leq h_{EAD}(S)\leq h_{EAR}(S)$
and $h_{SR}(S)\leq h_{AD}(S)\leq h_{AR}(S)$. One can construct an o-decision
tree $\Gamma $ over $S$, which solves the problem $EAR(S)$ by sequential
computation of values of all attributes from $A(S)$. Therefore $%
h_{EAR}(S)\leq n(S)$.
\end{proof}

Let $S$ be a decision rule system, $\Gamma $ be a decision tree over $S$,
and $\alpha =\{a_{i_{1}}=\delta _{1},\ldots ,a_{i_{m}}=\delta _{m}\}$ be a
consistent equation system such that $a_{i_{1}},\ldots ,a_{i_{m}}\in A$ and $%
\delta _{1},\ldots ,\delta _{m}\in \omega \cup \{\ast \}$. We now define a
decision rule system $S_{\alpha }$. Let $r$ be a decision rule for which the
equation system $K(r)\cup \alpha $ is consistent. We denote by $r_{\alpha }$
the decision rule obtained from $r$ by the removal from the left-hand side
of $r$ all equations that belong to $\alpha $. Then $S_{\alpha }$ is the set
of decision rules $r_{\alpha }$ such that $r\in S$ and the equation system $%
K(r)\cup \alpha $ is consistent.

Let us assume that $n(S)>0$,  $a_{i_{1}},\ldots ,a_{i_{m}}\in A(S)$, $\delta _{j}\in EV_{S}(a_{i_{j}})$ for $j=1,\ldots ,m$
and $\Gamma $ is an e-tree. We now define an e-tree $\Gamma _{\alpha }$.
First, we define a tree $\Gamma _{\alpha }^{\prime }$. Let an edge $d$ in
the tree $\Gamma $ enter a node $v$. We will say that a subtree of $\Gamma $
with the root $v$ \emph{corresponds to the edge} $d$. Denote $A(\alpha
)=\{a_{i_{1}},\ldots ,a_{i_{m}}\}$. Beginning with the root of $\Gamma $, we
will process all working nodes of the tree $\Gamma $. Let $v$ be a working
node of $\Gamma $ labeled with an attribute $a_{t}$. Let $a_{t}\in
A(S_{\alpha })$. We keep all edges that leave $v$ and are labeled with
elements from the set $EV_{S_{\alpha }}(a_{t})$. We remove all other edges
leaving $v$ together with the subtrees corresponding to these edges. Let $%
a_{t}\notin A(S_{\alpha })$. We choose an element $\delta \in EV_{S}(a_{t})$
in the following way. If $a_{t}\in A(\alpha )$ and $t=i_{j}$, then $\delta
=\delta _{j}$. If $a_{t}\notin A(\alpha )$, then $\delta $ is the minimum
number from $EV_{S}(a_{t})$. We remove all edges that leave $v$ and are
labeled with elements different from $\delta $ together with the subtrees
corresponding to these edges. Denote by $\Gamma _{\alpha }^{\prime }$ the
obtained tree.

We now process all nodes in this tree. Let $v$ be a working node that is
labeled with an attribute $a_{t}$. If $a_{t}\in A(S_{\alpha })$, then we
keep this node untouched. Let $a_{t}\notin A(S_{\alpha })$. If $v$ is the
root, then remove the node $v$ and the edge leaving $v$. Let $v$ be not the
root, $d_{0}$ be the edge entering $v$, and $d_{1}$ be the edge leaving $v$
and entering a node $v_{1}$. We remove from $\Gamma _{\alpha }^{\prime }$
the node $v$ and the edge $d_{1}$, and join the edge $d_{0}$ to the node $%
v_{1}$. Let $v$ be a terminal node that is labeled with a set $Z$ of
decision rules. We replace the set $Z$ with the set $Z_{\alpha }$. We denote
the obtained tree $\Gamma _{\alpha }$.

Let us assume that  $n(S)>0$,  $a_{i_{1}},\ldots ,a_{i_{m}}\in A(S)$, $\delta _{j}\in V_{S}(a_{i_{j}})$ for $j=1,\ldots ,m$ and
$\Gamma $ be an o-tree. The tree $\Gamma _{\alpha }^{\prime }$ and the
o-tree $\Gamma _{\alpha }$ are defined in almost the same way as in the
previous case. The only difference is that instead of the sets $%
EV_{S}(a_{t}) $ and $EV_{S_{\alpha }}(a_{t})$ we consider the sets $%
V_{S}(a_{t})$ and $V_{S_{\alpha }}(a_{t})$.

\begin{exmp}
Let us consider a decision rule system $S=\{r^1:(a_{1}=0)\wedge (a_{2}=0)\rightarrow 0, r^2 : (a_{1}=1)\wedge (a_{3}=0)\rightarrow 0, r^3 : (a_{2}=1)\wedge (a_{3}=1)\rightarrow 0\}$ and $\alpha =\{a_{3}=0\}$. Then $S_{\alpha } = \{r^1_{\alpha}:(a_{1}=0)\wedge (a_{2}=0)\rightarrow 0, r^2_{\alpha} :(a_{1}=1)\rightarrow 0\}$. In Fig. \ref{fig:fig3}, decision trees  $\Gamma$, $\Gamma _{\alpha }^{\prime }$, and $\Gamma _{\alpha }$ are depicted, where $\Gamma$ is an o-decision tree over $S$.

\begin{figure}[h!]
\renewcommand{\figurename}{\textbf{Fig.}}
    \centering
    \includegraphics[width=0.7\textwidth]{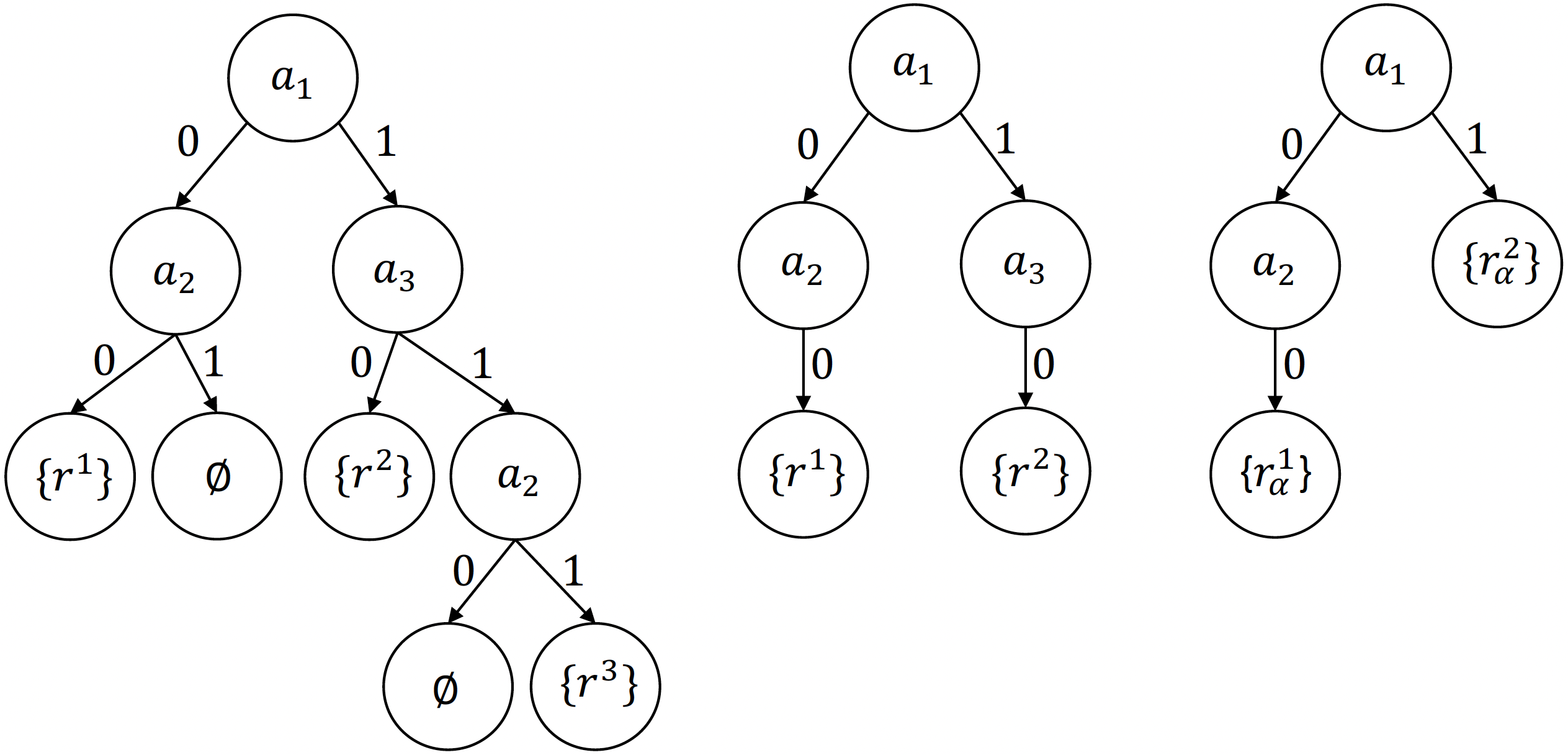}
    \caption{Decision trees $\Gamma$, $\Gamma _{\alpha }^{\prime }$, and $\Gamma _{\alpha }$}
    \label{fig:fig3}
\end{figure}

\end{exmp}

\begin{lemma}
\label{L4} Let $S$ be a decision rule system with $n(S)>0$, $\Gamma $ be a decision tree
over $S$, $C\in \{EAR,EAD,$ $ESR\}$, and $\alpha =\{a_{i_{1}}=\delta
_{1},\ldots ,a_{i_{m}}=\delta _{m}\}$ be a consistent equation system such
that $a_{i_{j}}\in A(S)$ and $\delta _{j}\in EV_{S}(a_{i_{j}})$ for $%
j=1,\ldots ,m$. If the decision tree $\Gamma $ solves the problem $C(S)$ and
$S_{\alpha }\neq \emptyset $, then the decision tree $\Gamma _{\alpha }$
solves the problem $C(S_{\alpha })$.
\end{lemma}

\begin{proof}
Let $\Gamma $ solve the problem $C(S)$ and $S_{\alpha }\neq \emptyset $. It
is clear that $\Gamma _{\alpha }$ is an e-decision tree over the decision
rule system $S_{\alpha }$. It is also clear that each complete path in $%
\Gamma _{\alpha }^{\prime }$ coincides with some complete path in $\Gamma $
and there exists one-to-one correspondence between sets of complete paths in
$\Gamma _{\alpha }^{\prime }$ and $\Gamma _{\alpha }$ such that each path $%
\xi $ in $\Gamma _{\alpha }^{\prime }$ corresponds to a path $\xi _{\alpha }$
in $\Gamma _{\alpha }$ obtained from $\xi $ by the removal of some nodes and
edges leaving these nodes. One can show that the equation system $K(\xi )$
is inconsistent if and only if the system $K(\xi _{\alpha })$ is
inconsistent.

Let $\xi $ be a complete path in $\Gamma _{\alpha }^{\prime }$ for which the
equation system $K(\xi )$ is consistent. Let the terminal node of the path $%
\xi $ be labeled with a set of decision rules $Z$. Then the terminal node of
the path $\xi _{\alpha }$ is labeled with the set of decision rules $%
Z_{\alpha }$. Let $r\in Z$. Since $\Gamma $ solves the problem $C(S)$, $%
K(r)\subseteq K(\xi )$. Therefore the system $K(r)\cup \alpha $ is
consistent. It is clear that $K(\xi _{\alpha })\subseteq K(\xi )$ and the
system $K(\xi )\setminus K(\xi _{\alpha })$ contains only equations of the
form $a_{t}=\delta $ such that $a_{t}\notin A(S_{\alpha })$. Since $%
K(r_{\alpha })\subseteq K(r)$, $K(r_{\alpha })\subseteq K(\xi _{\alpha })$.
Using these relations, we obtain $Z_{\alpha }=\{r_{\alpha }:r\in Z\}$ and $%
K(r_{\alpha })\subseteq K(\xi _{\alpha })$ for any $r_{\alpha }\in Z_{\alpha
}$.

Let $r\in S$, the system $K(r)\cup \alpha $ be consistent, and the system $%
K(r)\cup K(\xi )$ be inconsistent. Then there exist $a_{t}$, $\delta $, $%
\sigma $ such that $\delta \neq \sigma $, $a_{t}=\delta \in K(r)$ and $%
a_{t}=\sigma \in K(\xi )$. It is clear that $a_{t}\notin \{a_{i_{1}},\ldots
,a_{i_{m}}\}$. Therefore $a_{t}=\delta \in K(r_{\alpha })$ and $a_{t}\in A(S_{\alpha })$. Thus, $a_{t}=\sigma
\in K(\xi _{\alpha })$ and $K(r_{\alpha })\cup K(\xi _{\alpha })$ is
inconsistent.

Let $\Gamma $ solve the problem $EAR(S)$. We know that $K(r_{\alpha
})\subseteq K(\xi _{\alpha })$ for any $r_{\alpha }\in Z_{\alpha}$. Let $%
r^{\prime }\in S_{\alpha }\setminus Z_{\alpha }$ and $r$ be a rule from $S$
such that the system $K(r)\cup \alpha $ is consistent and $r^{\prime
}=r_{\alpha }$. It is clear that $r\notin Z$. Taking into account that $%
\Gamma $ solves the problem $EAR(S)$, we obtain that the system $K(r)\cup
K(\xi )$ is inconsistent. Therefore the system $K(r_{\alpha })\cup K(\xi
_{\alpha })$ is inconsistent. Hence $\Gamma _{\alpha }$ solves the problem $%
EAR(S_{\alpha })$.

Let $\Gamma $ solve the problem $EAD(S)$. We know that $K(r_{\alpha
})\subseteq K(\xi _{\alpha })$ for any $r_{\alpha }\in Z_{\alpha}$. Let $%
r^{\prime }\in S_{\alpha }\setminus Z_{\alpha }$ and the right-hand side of
the rule $r^{\prime }$ do not belong to $D(Z_{\alpha })$. We now consider a
decision rule $r\in S$ such that the system $K(r)\cup \alpha $ is consistent
and $r^{\prime }=r_{\alpha }$. It is clear that $r\notin Z$ and $%
D(Z)=D(Z_{\alpha })$. Taking into account that $\Gamma $ solves the problem $%
EAD(S)$, we obtain that the system $K(r)\cup K(\xi )$ is inconsistent.
Therefore the system $K(r_{\alpha })\cup K(\xi _{\alpha })$ is inconsistent.
Hence $\Gamma _{\alpha }$ solves the problem $EAD(S_{\alpha })$.

Let $\Gamma $ solve the problem $ESR(S)$. We know that $K(r_{\alpha
})\subseteq K(\xi _{\alpha })$ for any $r_{\alpha }\in Z_{\alpha}$. Let $%
Z_{\alpha }=\emptyset $. Then $Z=\emptyset $. Taking into account that $%
\Gamma $ solves the problem $ESR(S) $, we obtain that the system $K(r)\cup
K(\xi )$ is inconsistent for any rule $r\in S$ such that the system $%
K(r)\cup \alpha $ is consistent. Therefore the system $K(r_{\alpha })\cup
K(\xi _{\alpha })$ is inconsistent for any rule $r_{\alpha } \in S_{\alpha }$%
. Hence $\Gamma _{\alpha }$ solves the problem $ESR(S_{\alpha })$
\end{proof}

Proof of the following lemma is similar to the proof of Lemma \ref{L4}.

\begin{lemma}
\label{L5} Let $S$ be a decision rule system with $n(S)>0$, $\Gamma $ be a decision tree
over $S$, $C\in \{AR,AD,SR\}$, $\alpha =\{a_{i_{1}}=\delta _{1},\ldots
,a_{i_{m}}=\delta _{m}\}$ be a consistent equation system such that $%
a_{i_{j}}\in A(S)$ and $\delta _{j}\in V_{S}(a_{i_{j}})$ for $j=1,\ldots ,m$%
. If the decision tree $\Gamma $ solves the problem $C(S)$ and $S_{\alpha
}\neq \emptyset $, then the decision tree $\Gamma _{\alpha }$ solves the
problem $C(S_{\alpha })$.
\end{lemma}

\begin{lemma}
\label{L6} Let $S$ be a decision rule system with $n(S)>0$, $C\in \{EAR,AR,EAD,AD,ESR,$ $SR\}$%
, $\alpha =\{a_{i_{1}}=\delta _{1},\ldots ,a_{i_{m}}=\delta _{m}\}$ be a
consistent equation system such that $a_{i_{1}},\ldots ,a_{i_{m}}\in A(S)$
and, for $j=1,\ldots ,m$,  $\delta _{j}\in
EV_{S}(a_{i_{j}})$ if $C\in \{EAR,EAD,ESR\}$ and $\delta _{j}\in
V_{S}(a_{i_{j}})$ if $C\in \{AR,AD,SR\}$. Then $h_{C}(S)\geq h_{C}(S_{\alpha })$.
\end{lemma}

\begin{proof}
Let $\Gamma $ be a decision tree over $S$, which solves the problem $C(S)$
and for which $h(\Gamma )=h_{C}(S)$. Using Lemmas \ref{L4} and \ref{L5}, we
obtain that the decision tree $\Gamma _{\alpha }$ solves the problem $C(S_{\alpha })$. It is
clear that $h(\Gamma _{\alpha })\leq h(\Gamma )$. Therefore $h_{C}(S_{\alpha
})\leq h_{C}(S)$.
\end{proof}

We correspond to a decision rule system $S$ a hypergraph $G(S)$ with the set of nodes $%
A(S)$ and the set of edges $\{A(r):r\in S\}$. A \emph{node cover} of the
hypergraph $G(S)$ is a subset $B$ of the set of nodes $A(S)$ such that $%
A(r)\cap B\neq \emptyset $ for any rule $r\in S$ such that $A(r)\neq
\emptyset $. If $A(S)=\emptyset $, then the empty set is the only node cover of the
hypergraph $G(S)$. Denote by $\beta (S)$ the minimum cardinality of a node
cover of the hypergraph $G(S)$.

\begin{exmp}
Let us consider a decision rule system $S=\{(a_{1}=0)\wedge (a_{2}=0)\rightarrow 0, (a_{1}=1)\wedge (a_{3}=0)\rightarrow 0, (a_{2}=1)\wedge (a_{3}=1)\rightarrow 0\}$. One can show, that the set $\{a_{1}, a_{2}\}$ is a node cover of the hypergraph $G(S)$ and $\beta (S) = 2$.
\end{exmp}

We define a subsystem $I_{SR}(S)$ of the system $S$ in the following way. If
$S$ does not contain rules of the length $0$, then $I_{SR}(S)=S$. Otherwise,
$I_{SR}(S)$ consists of all rules from $S$ of the length $0$.

\begin{remark}
\label{R1}Note that $I_{SR}(S)\neq S$ if and only if $S$ contains both a
rule of the length $0$ and a rule of the length greater than $0$.
\end{remark}

We now define a subsystem $I_{AD}(S)$ of the system $S$. Denote by $D_{0}(S)$
the set of the right-hand sides of decision rules from $S$, which length is
equal to $0$. Then the subsystem $I_{AD}(S)$ consists of all rules from $S$
of the length $0$ and all rules from $S$ for which the right-hand sides do
not belong to $D_{0}(S)$.

\begin{remark}
\label{R2}Note that $I_{AD}(S)\neq S$ if and only if $S$ contains both a
rule of the length $0$ and a rule of the length greater than $0$ with the
same right-hand sides.
\end{remark}

\begin{exmp}
Let us consider a decision rule system $S=\{(a_{1}=0)\rightarrow 0, (a_{1}=1)\rightarrow 1, \rightarrow 1, \rightarrow 2\}$. For this system, $I_{SR}(S)=\{\rightarrow 1, \rightarrow 2\}$ and $I_{AD}(S)=\{(a_{1}=0)\rightarrow 0, \rightarrow 1, \rightarrow 2\}$.
\end{exmp}

Let $S$ be a decision rule system. This system will be called \emph{%
incomplete} if there exists a tuple $\bar{\delta}\in V(S)$ such that the
equation system $K(r)\cup K(S,\bar{\delta})$ is inconsistent for any
decision rule $r\in S$. Otherwise, the system $S$ will be called \emph{complete}. If $n(S)=0$, then the system $S$ will be considered as complete.

\begin{lemma}
\label{L7} Let $S$ be a decision rule system. Then

(a) If $C\in \{EAR,AR\}$, then $h_{C}(S)\geq \beta (S)$.

(b) If $C\in \{AD,SR\}$, then $h_{EC}(S)\geq \beta (I_{C}(S))$.

(c) If $C\in \{AD,SR\}$ and the system $S$ is incomplete, then $h_{C}(S)\geq
\beta (S)$.
\end{lemma}

\begin{proof}
It is clear that the statements of the lemma hold if $n(S)=0$. Let $n(S)>0$.

(a) Let $\Gamma $ be a decision tree over $S$, which solves the problem $%
AR(S)$ and for which $h(\Gamma )=h_{AR}(S)$. Let $\xi $ be a complete path
in $\Gamma $ for which the equation system $K(\xi )$ is consistent. It is
clear that, for any decision rule $r\in S$, either $K(r)\subseteq K(\xi )$
or the equation system $K(r)\cup K(\xi )$ is inconsistent. Hence $A(\xi
)\cap A(r)\neq \emptyset $ if $ A(r) \neq \emptyset$ and $A(\xi )$ is a node cover of the hypergraph $%
G(S)$. Thus, $h(\Gamma )\geq h(\xi )\geq \left\vert A(\xi )\right\vert \geq
\beta (S)$ and $h_{AR}(S)\geq \beta (S)$. Using Lemma \ref{L3}, we obtain $%
h_{EAR}(S)\geq \beta (S)$.

(b) If $S$ contains rules of the length $0$, then $d(I_{SR}(S))=0$, $\beta
(I_{SR}(S))=0$ and the inequality $h_{ESR}(S)\geq \beta (I_{SR}(S))$ holds.
Let $S$ does not contain rules of the length $0$. Then $I_{SR}(S)=S$. Let $%
\Gamma $ be a decision tree over $S$, which solves the problem $ESR(S)$ and
for which $h(\Gamma )=h_{ESR}(S)$. Let $\bar{\delta}=(\ast ,\ldots ,\ast
)\in EV(S)$ and $\xi $ be a complete path in $\Gamma $ such that $K(\xi
)\subseteq K(S,\bar{\delta})$. It is clear that the terminal node of this
path is labeled with the empty set of decision rules. Therefore, for any $%
r\in S$, the equation system $K(\xi )\cup K(r)$ is inconsistent. Hence $%
A(\xi )\cap A(r)\neq \emptyset $. Thus, $A(\xi )$ is a node cover of the
hypergraph $G(S)$. It is clear that $h(\Gamma )\geq \left\vert A(\xi
)\right\vert $ and $\left\vert A(\xi )\right\vert \geq \beta (S)$. Therefore
$h(\Gamma )\geq \beta (S)$ and $h_{ESR}(S)\geq \beta (S)=\beta (I_{SR}(S))$.

Let $\Gamma $ be a decision tree over $S$, which solves the problem $EAD(S)$
and for which $h(\Gamma )=h_{EAD}(S)$. Let $\bar{\delta}=(\ast ,\ldots ,\ast
)\in EV(S)$ and $\xi $ be a complete path in $\Gamma $ such that $K(\xi
)\subseteq K(S,\bar{\delta})$. It is clear that $\tau (\xi )$ contains only
rules of the length $0$ and $D(\tau (\xi ))=D_{0}(S)$. It is clear also
that, for any rule $r\in S$ such that the right-hand side of $r$ does not
belong to the set $D_{0}(\tau (\xi ))$, the system of equations $K(r)\cup
K(\xi )$ is inconsistent. Therefore $A(\xi )$ is a node cover of the
hypergraph $G(I_{AD}(S))$. It is clear that $h(\Gamma )\geq \left\vert A(\xi
)\right\vert $ and $\left\vert A(\xi )\right\vert \geq \beta (I_{AD}(S))$.
Therefore $h(\Gamma )\geq \beta (I_{AD}(S))$ and $h_{EAD}(S)\geq \beta
(I_{AD}(S))$.

(c) Let $\Gamma $ be a decision tree over $S$, which solves the problem $%
SR(S)$ and for which $h(\Gamma )=h_{SR}(S)$. Let the decision rule system $S$
be incomplete and $\bar{\delta}$ be a tuple from $V(S)$ for which the
equation system $K(r)\cup K(S,\bar{\delta})$ is inconsistent for any
decision rule $r\in S$. Let us consider a complete path $\xi$ in $\Gamma $
such that $K(\xi )\subseteq K(S,\bar{\delta})$. It is clear that the
terminal node of this path is labeled with the empty set of decision rules.
Therefore, for any rule $r\in S$, the equation system $K(\xi )\cup K(r)$ is
inconsistent. Hence $A(\xi )$ is a node cover of the hypergraph $G(S)$.
Therefore $h(\Gamma )\geq h(\xi )\geq \left\vert A(\xi )\right\vert \geq
\beta (S)$ and $h_{SR}(S)\geq \beta (S)$. Using Lemma \ref{L3}, we obtain $%
h_{AD}(S)\geq \beta (S)$.
\end{proof}

Note that the condition of incompleteness in the statement (c) of Lemma \ref%
{L7} is essential. Let us consider a decision rule system $%
S=\{(a_{1}=0)\rightarrow 0,\ldots ,(a_{n}=0)\rightarrow 0\}$. It is easy to
see that this system is complete and $\beta (S)=n$. We now consider a
decision tree $\Gamma $ that contains two nodes $v_{1}$, $v_{2}$ and an edge
$d$ leaving the node $v_{1}$ and entering the node $v_{2}$. The node $v_{1}$
is labeled with the attribute $a_{1}$, the node $v_{2}$ is labeled with the
decision rule system $\{(a_{1}=0)\rightarrow 0\}$, and the edge $d$ is
labeled with the number $0$. One can show that $\Gamma $ solves the problems
$AD(S)$ and $SR(S)$, and $h(\Gamma )=1$.

Note also that the incompleteness is not a hereditary property: there exists
an incomplete decision rule system $S$ and a consistent system of equations $%
\alpha =\{a_{i_{1}}=\delta _{1},\ldots ,a_{i_{m}}=\delta _{m}\}$ with $%
a_{i_{1}},\ldots ,a_{i_{m}}\in A(S)$ and $\delta _{1}\in
V_{S}(a_{i_{1}}),\ldots ,\delta _{m}\in V_{S}(a_{i_{m}})$ such that the decision
rule system $S_{\alpha }$ is complete. Let us consider a decision rule
system $S=\{(a_{1}=0)\wedge (a_{2}=0)\rightarrow 0,(a_{1}=1)\wedge
(a_{2}=0)\rightarrow 0,(a_{1}=0)\wedge (a_{2}=1)\rightarrow 0\}$. This
system is incomplete: for the tuple $(1,1)\in V(S)$, there is no decision
rule from $S$ that is realizable for this tuple. Let $\alpha =\{a_{2}=0\}$.
Then $S_{\alpha }=\{(a_{1}=0)\rightarrow 0,(a_{1}=1)\rightarrow 0\}$. It is
clear that the decision rule system $S_{\alpha }$ is complete.

\begin{lemma}
\label{L8} Let $S$ be a decision rule system. Then

(a) $h_{EAR}(S)\geq h_{AR}(S)\geq d(S)$.

(b) If $S$ is an $SR$-reduced system, then $h_{ESR}(S)\geq d(S)$.

(c) If $S$ is an $AD$-reduced system, then $h_{EAD}(S)\geq d(S)$.
\end{lemma}

\begin{proof}
It is clear that the statements of the lemma hold if $n(S)=0$. Let $n(S)>0$.

(a) Let $r$ be a decision rule from $S$ for which the length is equal to $%
d(S)$. Let $\Gamma $ be a decision tree over $S$, which solves the problem $%
AR(S)$ and for which $h(\Gamma )=h_{AR}(S)$. It is clear that there exists a
complete path $\xi $ in $\Gamma $ such that the equation system $K(r)\cup
K(\xi )$ is consistent. Taking into account that $\Gamma $ solves the
problem $AR(S)$, we obtain $K(r)\subseteq K(\xi )$. Therefore $h(\xi
)\geq d(S)$, $h(\Gamma )\geq d(S)$, and $h_{AR}(S)\geq d(S)$. Using Lemma %
\ref{L3}, we obtain $h_{EAR}(S)\geq h_{AR}(S)\geq d(S)$.

(b) Let $S$ be an $SR$-reduced system of decision rules, $r$ be a decision
rule of the length $d(S)$ from $S$, and $\Gamma $ be a decision tree over $S$%
, which solves the problem $ESR(S)$ and for which $h(\Gamma )=h_{ESR}(\Gamma
)$. We denote by $S(r)$ the set of rules $\rho$ from $S$ such that $K(\rho) = K(r)$. Let us consider a tuple $\bar{\delta}\in EV(S)$ such that $K(r)\subseteq
K(S,\bar{\delta})$ and each equation from the system $K(S,\bar{\delta}%
)\setminus K(r)$ has the form $a_{t}=\ast $. Taking into account that $S$ is
an $SR$-reduced system, we obtain that, for any rule $r^{\prime }\in
S\setminus S(r)$, the equation system $K(r^{\prime })\cup K(S,\bar{\delta})$
is inconsistent. Let $\xi $ be a complete path in $\Gamma $ such that $K(\xi
)\subseteq K(S,\bar{\delta})$. Since $\Gamma $ solves the problem $ESR(S)$,
the terminal node of the path $\xi $ is labeled with a nonempty subset of the set $S(r)$ and $%
K(\rho)\subseteq K(\xi )$ for any $\rho \in S(r)$. Therefore $h(\xi )\geq d(S)$, $h(\Gamma )\geq d(S)$,
and $h_{ESR}(S)\geq d(S)$.

(c) Let $S$ be an $AD$-reduced system of decision rules, $r$ be a decision
rule of the length $d(S)$ from $S$, and $\Gamma $ be a decision tree over $S$%
, which solves the problem $EAD(S)$ and for which $h(\Gamma )=h_{EAD}(\Gamma
)$. We denote by $S^{\prime }(r)$ the set of rules $\rho$ from $S$ such that $\rho$ and $r$ are equal. Let us consider a tuple $\bar{\delta}\in EV(S)$ such that $K(r)\subseteq
K(S,\bar{\delta})$ and each equation from the system $K(S,\bar{\delta}%
)\setminus K(r)$ has the form $a_{t}=\ast $. Taking into account that $S$ is
an $AD$-reduced system, we obtain that, for any rule $r^{\prime }\in
S\setminus S^{\prime }(r)$ such that $K(r^{\prime })\subseteq K(S,\bar{\delta})$, the
right-hand side of the rule $r^{\prime }$ is different from the right-hand
side of the rule $r$. Let $\xi $ be a complete path in $\Gamma $ such that $%
K(\xi )\subseteq K(S,\bar{\delta})$. Taking into account that $\Gamma $
solves the problem $EAD(S)$, one can show that at least one rule $r\in S^{\prime }(r)$ belongs to the
set $\tau (\xi )$ and $K(r)\subseteq K(\xi )$. Therefore $h(\xi )\geq d(S)$,
$h(\Gamma )\geq d(S)$, and $h_{EAD}(S)\geq d(S)$.
\end{proof}

Note that we cannot obtain for the problems $SR(S)$ and $AD(S)$ bounds similar to
those mentioned in statements (b) and (c) of Lemma \ref{L8}. Let us consider a
decision rule system $S=\{(a_{1}=0)\rightarrow 0,(a_{2}=0)\wedge \cdots
\wedge (a_{n}=0)\rightarrow 0\}$. It is easy to see that the system $S$ is $SR$-reduced
and $AD$-reduced, and $d(S)=n-1$. We now consider a decision tree $%
\Gamma $ that contains two nodes $v_{1}$, $v_{2}$ and an edge $d$ leaving
the node $v_{1}$ and entering the node $v_{2}$. The node $v_{1}$ is labeled
with the attribute $a_{1}$, the node $v_{2}$ is labeled with the decision
rule system $\{(a_{1}=0)\rightarrow 0\}$, and the edge $d$ is labeled with
the number $0$. One can show that $\Gamma $ solves the problems $AD(S)$ and $%
SR(S)$, and $h(\Gamma )=1$.

\begin{lemma}
\label{L9} Let $S$ be a decision rule system. Then $%
h_{ESR}(S)=h_{ESR}(R_{SR}(S))$ and $h_{EAD}(S)=h_{EAD}(R_{AD}(S))$.
\end{lemma}

\begin{proof}
It is clear that the considered equalities hold if $n(S)=0$. Let $n(S)>0$.

 Let us first show that $h_{ESR}(S)\leq h_{ESR}(R_{SR}(S))$.
Let $\Gamma $ be a decision tree over $R_{SR}(S)$, which solves the problem $%
ESR(R_{SR}(S))$ and for which $h(\Gamma )=h_{ESR}(R_{SR}(S))$. We will
process all nodes of the tree $\Gamma $ beginning with the terminal ones. If
$v$ is a terminal node of the tree $\Gamma $, then we will keep it
untouched. Let $v$ be a working node of the tree $\Gamma $ and let, for each
edge $d$ leaving $v$, all nodes in the subtree of $\Gamma $ corresponding to
$d$ be already processed. Let $v$ be labeled with an attribute $a_{i}$. If $%
EV_{S}(a_{i})=EV_{R_{SR}(S)}(a_{i})$, then we keep the node $v$ untouched.
Otherwise, for each $\delta \in EV_{S}(a_{i})\setminus EV_{R_{SR}(S)}(a_{i})$%
, we add to the tree $\Gamma $ a subtree $G$ corresponding to the edge that
leaves $v$ and is labeled with the symbol $\ast $. We also add an edge that
leaves $v$, enters the root of $G$, and is labeled with the number $\delta $%
. We denote by $\Gamma ^{\prime }$ the tree obtained after processing all
nodes of $\Gamma $. One can show that $\Gamma ^{\prime }$ is a decision tree
over $S$ that solves the problem $ESR(S)$. It is clear that $h(\Gamma
^{\prime })=h(\Gamma )$. Therefore $h_{ESR}(S)\leq h_{ESR}(R_{SR}(S))$.

We now show that $h_{ESR}(S)\geq h_{ESR}(R_{SR}(S))$. Let $\Gamma $ be a
decision tree over $S$, which solves the problem $ESR(S)$ and for which $h(\Gamma
)=h_{ESR}(S)$. We will process all nodes of the tree $\Gamma $ beginning
with the terminal ones. Let $v$ be a terminal node of $\Gamma $ and $v$ be
labeled with a set of decision rules $B$. If $B = \emptyset$, then we will keep the node $v$
untouched. Otherwise, replace each rule $r$ from $B$
with a rule $r^{\prime }\in R_{SR}(S)$ such that $K(r^{\prime })\subseteq
K(r)$. Let $v$ be a working node labeled with an attribute $a_{i}$ and, for
any edge $d$ leaving $v$, all nodes in the subtree of $\Gamma $
corresponding to $d$ be already processed. Let $a_{i}\notin A(R_{SR}(S))$
and $G$ be a subtree of $\Gamma $ corresponding to the edge $d$ that leaves $%
v$ and is labeled with the symbol $\ast $. If $v$ is the root of $\Gamma $,
then remove from $\Gamma $ all nodes and edges with the exception of the
subtree $G$. Let $v$ be not the root and $d^{\prime }$ be an edge that
enters $v$. We remove from the subtree corresponding to $d^{\prime }$  all nodes and edges with the exception of the
subtree $G$  and join $%
d^{\prime }$ to the root of $G$. Let $a_{i}$ $\in A(R_{SR}(S))$. We process
all edges leaving $v$. Let an edge $d$ leave $v$ and be labeled with a
number $\delta $. If $\delta \in EV_{R_{SR}(S)}(a_{i})$, then we will keep $%
d $ untouched. If $\delta \notin EV_{R_{SR}(S)}(a_{i})$, then we remove $d$
and corresponding to it subtree. We denote by $\Gamma ^{\prime }$ the tree
obtained after processing all nodes of $\Gamma $. One can show that $\Gamma
^{\prime }$ is a decision tree over $R_{SR}(S)$, which solves the problem $%
ESR(R_{SR}(S))$. It is clear that $h(\Gamma ^{\prime })\leq h(\Gamma )$.
Therefore $h_{ESR}(R_{SR}(S))\leq h_{ESR}(S)$. Thus, $%
h_{ESR}(R_{SR}(S))=h_{ESR}(S)$.

The second part of the lemma statement, the equality $%
h_{EAD}(S)=h_{EAD}(R_{AD}(S))$, can be proved similarly.
\end{proof}

\section{Unimprovable Upper Bounds}

\label{S4}

In this section, we study unimprovable upper bounds on the minimum depth of decision trees depending on three parameters of the decision rule systems.

First, we prove several lemmas.

\begin{lemma}
\label{L10}Let $n,d,k\in \omega \setminus \{0\}$ and $d\leq n$. Then the
following inequalities hold:

\begin{equation*}
\begin{array}{ccccccc}
H_{ESR}(n,d,k) & \leq & H_{EAD}(n,d,k) & \leq & H_{EAR}(n,d,k) & \leq & n \\
\mathrel{\rotatebox{90}{$\le$}} &  & \mathrel{\rotatebox{90}{$\le$}} &  & %
\mathrel{\rotatebox{90}{$\le$}} &  &  \\
H_{SR}(n,d,k) & \leq & H_{AD}(n,d,k) & \leq & H_{AR}(n,d,k), &  &
\end{array}%
\end{equation*}
\begin{equation*}
H_{SR}^{R}(n,d,k)\leq H_{ESR}^{R}(n,d,k)\leq n, H_{AD}^{R}(n,d,k)\leq
H_{EAD}^{R}(n,d,k)\leq n.
\end{equation*}
\end{lemma}

\begin{proof}
The considered inequalities follow immediately from Lemma \ref{L3}.
\end{proof}

\begin{lemma}
\label{L11}Let $n,d,k\in \omega \setminus \{0\}$ and $d\leq n$. Then the
following inequalities hold:
\begin{eqnarray*}
H_{SR}(n,d,k)\geq H_{SR}^{R}(n,d,k), H_{ESR}(n,d,k)\geq H_{ESR}^{R}(n,d,k),
\\
H_{AD}(n,d,k)\geq H_{AD}^{R}(n,d,k) , H_{EAD}(n,d,k)\geq H_{EAD}^{R}(n,d,k).
\end{eqnarray*}
\end{lemma}

\begin{proof}
The considered inequalities follow from the obvious inclusions $\Sigma
_{SR}\subset \Sigma $ and $\Sigma _{AD}\subset \Sigma $.
\end{proof}

\begin{lemma}
\label{L12}Let $n,d,k\in \omega \setminus \{0\}$ and $d\leq n$. Then
\begin{equation*}
H_{SR}^{R}(n,d,k)=H_{SR}(n,d,k)=\left\{
\begin{array}{cc}
1\text{,} & \text{if }d=1, \\
d\text{,} & \text{if }k=1, \\
n\text{,} & \text{if }k>1\text{ and }d>1.%
\end{array}%
\right.
\end{equation*}
\end{lemma}

\begin{proof}
Let $S\in \Sigma $ and $d(S)=1$. Let $a_{i}\in A(S)$. It is clear that, for
any $\delta \in V_{S}(a_{i})$, the system $S$ contains a rule $r_{\delta }$
of the form $(a_{i}=\delta )\rightarrow \sigma $. Let $V_{S}(a_{i})=\{\delta
_{1},\ldots ,\delta _{m}\}$. We denote by $\Gamma $ a decision tree over $S$
that consists of a node $v_{0}$ labeled with the attribute $a_{i}$, a node $%
v_{j}$ labeled with the set $\{r_{\delta _{j}}\}$, and an edge $d_{j}$,
which leaves the node $v_{0}$, enters the node $v_{j}$, and is labeled with
the number $\delta _{j}$, $j=1,\ldots ,m\,$. One can show that $\Gamma $
solves the problem $SR(S)$ and $h(\Gamma )=1$. Therefore $h_{SR}(S)\leq 1$.
Using Lemma \ref{L11}, we obtain $H_{SR}^{R}(n,1,k)\leq H_{SR}(n,1,k)\leq 1$%
. Let us consider a decision rule system $S=\{(a_{1}=0)\rightarrow 0,\ldots
,(a_{1}=k-1)\rightarrow 0,(a_{2}=0)\rightarrow 0,(a_{3}=0)\rightarrow
0,\ldots ,(a_{n}=0)\rightarrow 0\}$. It is clear that $S\in \Sigma _{SR}$, $%
n(S)=n$, $d(S)=1$, $k(S)=k$, and $h_{SR}(S)\geq 1$. Therefore $%
H_{SR}^{R}(n,1,k)\geq 1$. Thus, $H_{SR}^{R}(n,1,k)=H_{SR}(n,1,k)=1$.

Let $S\in \Sigma $ and $k(S)=1$. One can show that the value $h_{SR}(S)$ is
equal to the minimum length of a rule from $S$. Using this fact and Lemma %
\ref{L11}, we obtain that $H_{SR}^{R}(n,d,1)\leq H_{SR}(n,d,1)\leq d$. Let
us consider a system of decision rules $S=\{(a_{i}=0)\wedge \cdots \wedge
(a_{i+d-1}=0)\rightarrow i:i=1,\ldots ,n-d+1\}$. It is clear that $S\in
\Sigma _{SR}$, $n(S)=n$, $d(S)=d$, $k(S)=1$, and $h_{SR}(S)=d$. Therefore $%
H_{SR}^{R}(n,d,1)\geq d$ and $H_{SR}^{R}(n,d,1)=H_{SR}(n,d,1)=d$.

Let $k>1$ and $d>1$. We now show that $H_{SR}^{R}(n,d,k)\geq n$. Let us
consider a decision rule system
\begin{eqnarray*}
S =\{(a_{1}=1)\wedge \cdots \wedge (a_{d}=1)\rightarrow 0,(a_{1}=1)\wedge
(a_{2}=0)\rightarrow 1,(a_{2}=1)\wedge (a_{3}=0)\rightarrow 2,\ldots , \\
(a_{n-1} =1)\wedge (a_{n}=0)\rightarrow n-1,(a_{n}=1)\wedge
(a_{1}=0)\rightarrow n, \\
(a_{1}=2)\rightarrow 0,\ldots ,(a_{1}=k-1)\rightarrow 0\}.
\end{eqnarray*}%
It is clear that $S\in \Sigma _{SR}$, $n(S)=n,$ $d(S)=d$, and $k(S)=k$.
Let $\Gamma $ be a decision tree, which solves the problem $SR(S)$ and for
which $h(\Gamma )=h_{SR}(S)$. Let $\bar{\delta}=(0,\ldots ,0)\in V(S)$ and $%
\xi $ be a complete path in $\Gamma $ such that $K(\xi )\subseteq K(S,\bar{%
\delta})$. It is clear that there are no rules from $S$ that are realizable
for $\bar{\delta}$. Therefore the terminal node of $\xi $ is labeled with
the empty set of decision rules and, for any rule $r\in S$, the system of
equations $K(r)\cup K(\xi )$ is inconsistent. Let us assume that, for some $%
i\in \{1,\ldots ,n\}$, the equation $a_{i}=0$ does not belong to $K(\xi )$.
Then the system of equations $K(r_{i})\cup K(\xi )$ is consistent, where $%
r_{i}$ is the rule from $S$ with the right-hand side equal to $i$, but this
is impossible. Therefore $h(\xi )\geq n$, $h(\Gamma )\geq n$, and $%
h_{SR}(S)\geq n$. Hence $H_{SR}^{R}(n,d,k)\geq $ $n$. Using Lemmas \ref{L10}
and \ref{L11}, we obtain $H_{SR}^{R}(n,d,k)=H_{SR}(n,d,k)=n$.
\end{proof}

\begin{lemma}
\label{L13}Let $n,d,k\in \omega \setminus \{0\}$ and $d\leq n$. Then $%
H_{ESR}^{R}(n,d,k)=n$ and $H_{AD}^{R}(n,d,k)=n$.
\end{lemma}

\begin{proof}
Let us consider a system of decision rules $S=S_{1}\cup S_{2}$, where $%
S_{1}=\{(a_{1}=0)\wedge \cdots \wedge (a_{d}=0)\rightarrow
0,(a_{d+1}=0)\rightarrow d+1,\ldots ,(a_{n}=0)\rightarrow n\}$ and $%
S_{2}=\{(a_{1}=1)\rightarrow n+1,\ldots ,(a_{1}=k-1)\rightarrow n+k-1\}$. If
$k=1$, then $S_{2}=\emptyset $. It is clear that $S\in \Sigma _{SR}\cap
\Sigma _{AD}$, $n(S)=n$, $d(S)=d$, and $k(S)=k$.

Let $\Gamma _{1}$ be a decision tree, which solves the problem $AD(S)$ and
for which $h(\Gamma _1)=h_{AD}(S)$. Let $\bar{\delta}=(0,\ldots ,0)\in V(S)$
and $\xi $ be a complete path in $\Gamma _{1}$ such that $K(\xi )\subseteq
K(S,\bar{\delta})$. Taking into account that $\Gamma _{1}$ solves the
problem $AD(S)$, one can show that the terminal node of this path is labeled
with the set $S_{1}$ and $K(r)\subseteq K(\xi )$ for any $r\in S_{1}$.
Therefore $h(\xi )\geq n$. Hence $h(\Gamma
_{1})\geq n$ and $h_{AD}(S)\geq n$. Thus, $H_{AD}^{R}(n,d,k)\geq n$. Using
Lemma \ref{L10}, we obtain that $H_{AD}^{R}(n,d,k)=n$.

Let $\Gamma _{2}$ be a decision tree, which solves the problem $ESR(S)$ and
for which $h(\Gamma _{2})=h_{ESR}(S)$. Let $\bar{\delta}=(\delta _{1},\ldots
,\delta _{n})$ be a tuple from $EV(S)$ such that $\delta _{1}=\cdots =\delta
_{d}=0$ and $\delta _{d+1}=\cdots =\delta _{n}=\ast $. Let $\xi $ be a
complete path in $\Gamma _{2}$ for which $K(\xi )\subseteq K(S,\bar{\delta})$%
. We denote by $r_{0}$ the decision rule $(a_{1}=0)\wedge \cdots \wedge
(a_{d}=0)\rightarrow 0$. It is clear that the only decision rule $r_{0}$
from $S$ is realizable for the tuple $\bar{\delta}$. Taking into account
that $\Gamma _{2}$ solves the problem $ESR(S)$, we obtain that $%
K(r_{0})\subseteq K(\xi )$. Let $\xi =v_{1},d_{1},\ldots
,v_{m},d_{m},v_{m+1} $. For $i=1,\ldots ,d$, let $j_{i}$ be the minimum
number from the set $\{1,\ldots ,m\}$ such that the node $v_{j_{i}}$ is
labeled with the attribute $a_{i}$. Let $i_{0}$ be a number from $\{1,\ldots
,d\}$ such that $j_{i_{0}}=\max \{j_{1},\ldots ,j_{d}\}$. Let us consider
the tuple $\bar{\sigma}=(\sigma _{1},\ldots ,\sigma _{n})$ from $EV(S)$ such
that $\sigma _{1}=\cdots =\sigma _{i_{0}-1}=\sigma _{i_{0}+1}=\cdots =\sigma
_{d}=0$ and $\sigma _{i_{0}}=\sigma _{d+1}=\cdots =\sigma _{n}=\ast $. Let $%
\xi ^{\prime } $ be a complete path in $\Gamma _{2}$ such that $K(\xi
^{\prime })\subseteq K(S,\bar{\sigma})$. It is clear that there are no
decision rules from $S$ that are realizable for the tuple $\bar{\sigma}$.
Taking into account that $\Gamma _{2}$ solves the problem $ESR(S)$, we
obtain that the system of equations $K(r)\cup K(\xi ^{\prime })$ is
inconsistent for any $r\in S$. It is clear that the paths $\xi $ and $\xi
^{\prime }$ have $j_{i_{0}}$ common nodes $v_{1},\ldots ,v_{j_{i_{0}}}$.
Therefore $a_{1},\ldots ,a_{d}\in A(\xi ^{\prime })$. Let us assume that $%
a_{j}\notin A(\xi ^{\prime })$ for some $j\in \{d+1,\ldots ,n\}$. Then the
system of equations $K((a_{j}=0)\rightarrow j)\cup K(\xi ^{\prime })$ is
consistent, which is impossible. Therefore $\left\vert A(\xi ^{\prime
})\right\vert =n$ and $h(\xi ^{\prime })\geq n$. Hence $h(\Gamma _{2})\geq n$%
, $h_{ESR}(S)\geq n$, and $H_{ESR}^{R}(n,d,k)\geq n$. Using Lemma \ref{L10},
we obtain $H_{ESR}^{R}(n,d,k)=n$.
\end{proof}

\begin{theorem}
\label{T1} Let $n,d,k\in \omega \setminus \{0\}$ and $d\leq n$. Then

\begin{equation*}
H_{SR}^{R}(n,d,k)=H_{SR}(n,d,k)=\left\{
\begin{array}{cc}
1\text{,} & \text{if }d=1, \\
d\text{,} & \text{if }k=1, \\
n\text{,} & \text{if }k>1\text{ and }d>1%
\end{array}%
\right.
\end{equation*}
and
\begin{eqnarray*}
H_{ESR}^{R}(n,d,k)
=H_{AD}^{R}(n,d,k)=H_{EAD}^{R}(n,d,k)=H_{AD}(n,d,k)=H_{AR}(n,d,k) \\
=H_{ESR}(n,d,k)=H_{EAD}(n,d,k)=H_{EAR}(n,d,k)=n.
\end{eqnarray*}
\end{theorem}

\begin{proof}
The first statement of the theorem follows from Lemma \ref{L12}. The second statement  follows from Lemmas \ref{L10}, \ref{L11}, and \ref{L13}.
\end{proof}

\begin{remark}
Note that the equality $H_{AR}(n,d,k)=n$ mentioned in Theorem \ref{T1} was
obtained in \cite{Moshkov98}. The equality $H_{EAR}(n,d,k)=n$ mentioned in Theorem \ref{T1} was published
in \cite{Moshkov01} without proof.
\end{remark}

\section{Unimprovable Lower Bounds}

\label{S5}

In this section, we study unimprovable  lower bounds on the minimum depth of decision trees depending on three parameters of the decision rule systems.

First, we consider two simple statements.

\begin{lemma}
\label{L14}Let $n,d,k\in \omega \setminus \{0\}$ and $d\leq n$. Then the
following inequalities hold:

\begin{equation*}
\begin{array}{ccccccc}
h_{ESR}(n,d,k) & \leq & h_{EAD}(n,d,k) & \leq & h_{EAR}(n,d,k) & \leq & n \\
\mathrel{\rotatebox{90}{$\le$}} &  & \mathrel{\rotatebox{90}{$\le$}} &  & %
\mathrel{\rotatebox{90}{$\le$}} &  &  \\
h_{SR}(n,d,k) & \leq & h_{AD}(n,d,k) & \leq & h_{AR}(n,d,k), &  &
\end{array}%
\end{equation*}

\begin{equation*}
h_{SR}^{R}(n,d,k)\leq h_{ESR}^{R}(n,d,k)\leq n, h_{AD}^{R}(n,d,k)\leq
h_{EAD}^{R}(n,d,k)\leq n.
\end{equation*}
\end{lemma}

\begin{proof}
The considered inequalities follow immediately from Lemma \ref{L3}.
\end{proof}

\begin{lemma}
\label{L15}Let $n,d,k\in \omega \setminus \{0\}$ and $d\leq n$. Then the
following inequalities hold:
\begin{eqnarray*}
h_{SR}(n,d,k)\leq h_{SR}^{R}(n,d,k), h_{ESR}(n,d,k)\leq h_{ESR}^{R}(n,d,k),
\\
h_{AD}(n,d,k)\leq h_{AD}^{R}(n,d,k), h_{EAD}(n,d,k)\leq h_{EAD}^{R}(n,d,k).
\end{eqnarray*}
\end{lemma}

\begin{proof}
The considered inequalities follow from the obvious inclusions $\Sigma
_{SR}\subset \Sigma $ and $\Sigma _{AD}\subset \Sigma $.
\end{proof}

\subsection{Bounds on $h_{SR}(n,d,k)$, $h_{AD}(n,d,k)$, $h_{ESR}(n,d,k)$,
$h_{EAD}(n,d,k)$, $h_{SR}^{R}(n,d,k)$, and $h_{AD}^{R}(n,d,k)$ \label%
{S5.1}}

\begin{lemma}
\label{L16}Let $n,d,k\in \omega \setminus \{0\}$ and $d\leq n$. Then
\begin{equation*}
h_{SR}(n,d,k)=h_{AD}(n,d,k)=h_{ESR}(n,d,k)=h_{EAD}(n,d,k)=0.
\end{equation*}
\end{lemma}

\begin{proof}
Let $C\in \{SR,AD,ESR,EAD\}$. We now consider a decision rule system
\begin{eqnarray*}
S =\{\rightarrow 0,(a_{1}=0)\wedge \cdots \wedge (a_{d}=0)\rightarrow
0,(a_{d+1}=0)\rightarrow 0,\ldots ,(a_{n}=0)\rightarrow 0, \\
(a_{1} =1)\rightarrow 0,\ldots ,(a_{1}=k-1)\rightarrow 0\}.
\end{eqnarray*}%
It is clear that $n(S)=n$, $d(S)=d$, and $k(S)=k$. Let $\Gamma $ be a
decision tree that consists of one node, which is labeled with the decision
rule set $\{\rightarrow 0\}$. It is easy to show that $\Gamma $ solves the problem $%
C(S)$. Therefore $h_{C}(n,d,k)\leq 0$. Evidently, $h_{C}(n,d,k)\geq 0$.
Thus, $h_{C}(n,d,k)=0$.
\end{proof}

\begin{lemma}
\label{L17} Let $n,d,k\in \omega \setminus \{0\}$ and $d\leq n$. Then%
\begin{equation*}
h_{SR}^{R}(n,d,k)=h_{AD}^{R}(n,d,k)=\left\{
\begin{array}{ll}
n, & \text{if }d=n, \\
1, & \text{if }d\neq n.%
\end{array}%
\right.
\end{equation*}
\end{lemma}

\begin{proof}
Let $S\in \Sigma _{SR}$, $n(S)=n$, $d(S)=n$, and $k(S)=k$. Let $%
A(S)=\{a_{i_{1}},\ldots ,a_{i_{n}}\}$. Then there exists a decision rule $%
r_{0}$ from $S$ of the form $(a_{i_{1}}=\delta _{1})\wedge \cdots \wedge
(a_{i_{n}}=\delta _{n})\rightarrow \sigma $. Let $\Gamma $ be a decision
tree, which solves the problem $SR(S)$ and for which $h(\Gamma )=h_{SR}(S)$.
Let $\bar{\delta}=(\delta _{1},\ldots ,\delta _{n})$ and $\xi $ be a
complete path in $\Gamma $ such that $K(\xi )\subseteq K(S,\bar{\delta})$.
Since $S$ is an $SR$-reduced system of decision rules, only decision
rules $r$ from $S$ with $K(r)= K(r_0 )$ are realizable for the tuple $\bar{\delta}$. Since $%
\Gamma $ solves the problem $SR(S)$, $K(r_{0})\subseteq K(\xi )$. Hence $%
h(\xi )\geq n$, $h(\Gamma )\geq n$, $h_{SR}(S)\geq n$, and $%
h_{SR}^{R}(n,n,k)\geq n$. Using Lemma \ref{L14}, we obtain $%
h_{SR}^{R}(n,n,k)=n$.

Let $S\in \Sigma _{AD}$, $n(S)=n$, $d(S)=n$, and $k(S)=k$. Let $%
A(S)=\{a_{i_{1}},\ldots ,a_{i_{n}}\}$. Then there exists a decision rule $%
r_{0}$ from $S$ of the form $(a_{i_{1}}=\delta _{1})\wedge \cdots \wedge
(a_{i_{n}}=\delta _{n})\rightarrow \sigma $. Let $\Gamma $ be a decision
tree, which solves the problem $AD(S)$ and for which $h(\Gamma )=h_{AD}(S)$.
Let $\bar{\delta}=(\delta _{1},\ldots ,\delta _{n})$ and $\xi $ be a
complete path in $\Gamma $ such that $K(\xi )\subseteq K(S,\bar{\delta})$.
It is clear that $r_{0}$ is realizable for the tuple $\bar{\delta}$. Taking
into account that $S$ is an $AD$-reduced system of decision rules, we obtain
that any rule $r\in S$, which is different from $r_{0}$ and is realizable
for the tuple $\bar{\delta}$, has the right-hand side different from $\sigma
$. Taking into account that $\Gamma $ solves the problem $AD(S)$, we obtain $%
K(r_{0})\subseteq K(\xi )$. Therefore $h(\Gamma )\geq n$, $h_{AD}(S)\geq n$,
and $h_{AD}^{R}(n,n,k)\geq n$. Using Lemma \ref{L14}, we obtain $%
h_{SR}^{R}(n,n,k)=n$.

Let $d<n$. Let us consider a decision rule system $S=S_{1}\cup S_{2}$, where
$S_{1}=\{(a_{1}=0)\wedge \cdots \wedge (a_{d}=0)\rightarrow
0,(a_{d+1}=0)\rightarrow 0,\ldots ,(a_{n}=0)\rightarrow 0\}$ and $%
S_{2}=\{(a_{1}=1)\rightarrow 0,\ldots ,(a_{1}=k-1)\rightarrow 0\}$. If $k=1$%
, then $S_{2}=\emptyset $. It is clear that $S\in \Sigma _{SR}\cap \Sigma
_{AD}$, $n(S)=n$, $d(S)=d$, and $k(S)=k$. Denote by $\Gamma $ a decision tree
that consists of nodes $v_{1}$, $v_{2}$ and edge $d$ leaving $v_{1}$ and
entering $v_{2}$. The node $v_{1}$ is labeled with the attribute $a_{n}$,
the node $v_{2}$ is labeled with the set of decision rules $\{(a_{n}=0)\rightarrow 0\}$, and
the edge $d$ is labeled with the number $0$. One can show that $h(\Gamma )=1$
and the decision tree $\Gamma $ solves the problems $SR(S)$ and $AD(S)$.
Therefore $h_{SR}^{R}(n,d,k)\leq 1$ and $h_{AD}^{R}(n,d,k)\leq 1$.

Let $S\in \Sigma _{SR}$ and $n(S)>0$. Then $S$ does not contain decision
rules of the length $0$. Using this fact, one can show that $h_{SR}(S)\geq 1$%
. Therefore $h_{SR}^{R}(n,d,k)=1$ if $d<n$.

Let $S\in \Sigma _{AD}$ and $n(S)>0$. Then $S$ contains a decision rule of
the form $(a_{i_{1}}=\delta _{1})\wedge \cdots \wedge (a_{i_{m}}=\delta
_{m})\rightarrow \sigma $, where $m>0$. Taking into account that $S\in
\Sigma _{AD}$, we obtain $S$ does not contain decision rules of the
form $\rightarrow \sigma $. Using these considerations, one can show that $%
h_{AD}(S)\geq 1$. Therefore $h_{AD}^{R}(n,d,k)=1$ if $d<n$.
\end{proof}

\subsection{Bounds on $h_{AR}(n,d,k)$ and $h_{EAR}(n,d,k)$ \label{S5.2}}

\begin{lemma}
\label{L18}Let $S$ be a decision rule system and $S^{\prime }$ be a nonempty
subsystem of the system $S$. Then $h_{AR}(S)\geq h_{AR}(S^{\prime })$.
\end{lemma}

\begin{proof}
It is clear that the considered inequality holds if $n(S)=0$. Let $n(S)>0$.

Let $\Gamma $ be a decision tree, which solves the problem $AR(S)$ and for
which $h(\Gamma )=h_{AR}(S)$. Starting from the root we will process all
working nodes of the tree $\Gamma $. Let $v$ be a working node labeled with
an attribute $a_{t}$. Let $a_{t}\in A(S^{\prime })$. We keep all edges
leaving $v$ that are labeled with numbers from the set $V_{S^{\prime
}}(a_{t})$. We remove all other edges leaving $v$ together with the subtrees
corresponding to these edges. Let $a_{t}\notin A(S^{\prime })$ and $\delta $
be the minimum number from $V_{S}(a_{t})$. We remove all edges leaving $v$
together with the subtrees corresponding to these edges with the exception
of the edge labeled with $\delta $. Denote by $\Gamma _{0}$ the tree
obtained from $\Gamma $ after processing of all working nodes.

We now process all nodes in the tree $\Gamma _{0}$. Let $v$ be a working
node labeled with an attribute $a_{t}$. If $a_{t}\in A(S^{\prime })$, then
leave the node $v$ untouched. Let $a_{t}\notin A(S^{\prime })$. If $v$ is
the root of the tree $\Gamma _{0}$, then remove the node $v$ and the edge leaving it. Let $%
v$ be not the root. Let $d_{0}$ be the edge entering $v$ and $d_{1}$ be the
edge leaving the node $v$. Let $d_{1}$ enter a node $v_{1}$. We remove from $%
\Gamma _{0}$ the node $v$ and the edge $d_{1}$, and join the edge $d_{0}$ to
the node $v_{1}$. Let $v$ be a terminal node labeled with a set of decision
rules $B$. Replace the set $B$ with the set $B\cap S^{\prime }$. Denote the
obtained tree $\Gamma ^{\prime }$.

One can show that $\Gamma ^{\prime }$ is an o-tree over the decision rule
system $S^{\prime }$. We now prove that $\Gamma ^{\prime }$ solves the
problem $AR(S^{\prime })$. Let $\xi ^{\prime }$ be a complete path in $%
\Gamma ^{\prime }$ with consistent system of equations $K(\xi ^{\prime })$. Then there exists a complete path $\xi $ in $\Gamma $
satisfying the following conditions: $\tau (\xi ^{\prime })=\tau (\xi )\cap
S^{\prime }$, $K(\xi ^{\prime })\subseteq K(\xi )$, and all equations from
the set $K(\xi )\setminus K(\xi ^{\prime })$ have the form $a_{t}=\delta $,
where $a_{t}\notin A(S^{\prime })$ and $\delta $
is the minimum number from the set $V_{S}(a_{t})$. It is easy to show that the  system of equations $K(\xi )$ is consistent.
Let $r\in \tau (\xi ^{\prime })$. Then $%
r\in \tau (\xi )$. Therefore $K(r)\subseteq K(\xi )$ and hence $%
K(r)\subseteq K(\xi ^{\prime })$. Let $r\in S^{\prime }\setminus \tau (\xi
^{\prime })$. Then $r\in S\setminus \tau (\xi )$. Thus, the system $K(r)\cup
K(\xi )$ is inconsistent. Therefore the system $K(r)\cup K(\xi ^{\prime })$
is also inconsistent. As a result, we obtain that $\Gamma ^{\prime }$ solves
the problem $AR(S^{\prime })$. It is clear that $h(\Gamma ^{\prime })\leq
h(\Gamma )$. Thus, $h_{AR}(S^{\prime })\leq h(\Gamma )=h_{AR}(S)$.
\end{proof}

\begin{lemma}
\label{L19} Let $n,d,k\in \omega \setminus \{0\}$ and $d\leq n$. Then

(a) If $k=1$ or $d=1$, then $h_{AR}(n,d,k)=n$.

(b) If $k\ge 2$ and $d\ge 2$, then $h_{AR}(n,d,k)\geq \max \left\{ d,\frac{n(k-1)}{%
k^{d}}\right\} $.
\end{lemma}

\begin{proof}
(a) Let $S$ be a decision rule system with $n(S)=n$, $d(S)=d$, and $k(S)=k$.

Let $k=1$ and $\Gamma $ be a decision tree, which solves the problem $AR(S)$
and for which $h(\Gamma )=h_{AR}(S)$. It is clear that the decision tree $%
\Gamma $ has exactly one complete path $\xi $. The terminal node of this
path is labeled with the set $S$ and $K(r)\subseteq K(\xi )$ for any rule $%
r\in S$. It is clear that $h(\xi )\geq n$, $h(\Gamma )\geq n$, and $%
h_{AR}(S)\geq n$. Taking into account that $S$ is an arbitrary decision rule
system with $k(S)=1$, we obtain $h_{AR}(n,d,1)\geq n$. By Lemma \ref{L14}, $%
h_{AR}(n,d,1)=n$.

Let $d=1$. One can show that in this case $\beta (S)=n$. By Lemma \ref{L7}, $%
h_{AR}(S)\geq n$. Taking into account that $S$ is an arbitrary decision rule
system with $d(S)=1$, we obtain $h_{AR}(n,1,k)\geq n$. From this inequality
and from Lemma \ref{L14} it follows that $h_{AR}(n,1,k)=n$.

(b) Let $S$ be a decision rule system. From Lemma \ref{L8} it follows that $%
h_{AR}(S)\geq d(S)$. Since $S$ is an arbitrary decision rule system, $%
h_{AR}(n,d,k)\geq d$.

We will prove by induction on $d$ that, for any $n,d,k\in \omega \setminus
\{0\}$ such that $d\leq n$, the following inequality holds:%
\begin{equation}
h_{AR}(n,d,k)\geq \frac{n(k-1)}{k^{d}}.  \label{E1}
\end{equation}%
From the equality $h_{AR}(n,1,k)=n$ proved above it follows that the
inequality (\ref{E1})  holds if $d=1$.
 Let us assume that, for some $d \ge 1$, the inequality (\ref{E1}) holds for any $d^{\prime} $, $1 \le d^{\prime} \le d $. We now show that (\ref{E1}) holds for $d+1$.
Let $S$ be a decision rule system with $n(S)=n$, $d(S)=d+1$, and $k(S)=k$.
We choose a subsystem $S^{\prime }$ of the system $S$ such that $A(S^{\prime
})=A(S)$ and $A(S_{1})\neq A(S)$ for any system $S_{1}\subset S^{\prime }$.
It is clear that $n(S^{\prime })=n$, $d(S^{\prime })\leq d+1$, and $%
k(S^{\prime })\leq k$. Using Lemma \ref{L18}, we obtain%
\begin{equation}
h_{AR}(S)\geq h_{AR}(S^{\prime }).  \label{E2}
\end{equation}%
Let $S^{\prime }=\{r_{1},\ldots ,r_{p}\}$. It is clear that, for $j=1,\ldots
,p$, the equation system $K(r_{j})$ contains an equation $a_{i_{j}}=\sigma _{j}$ such
that $a_{i_{j}}\notin A(S^{\prime })\setminus A(r_{j})$. Therefore
\begin{equation}
p=\left\vert S^{\prime }\right\vert \leq n.  \label{E3}
\end{equation}

Let us assume that $\left\vert S^{\prime }\right\vert \leq \frac{n(k-1)}{k}$%
. Denote $\alpha =\{a_{i_{1}}=\sigma _{1},\ldots ,a_{i_{p}}=\sigma _{p}\}$. It is clear that $\alpha$ is a consistent equation system.
We now consider the system of decision rules $S_{\alpha }^{\prime }$
(corresponding definition is before Lemma \ref{L4}). Denote $n_0=n(S_{\alpha }^{\prime })$, $d_0=d(S_{\alpha }^{\prime })$, and $k_0=k(S_{\alpha }^{\prime })$.
One can show that $n_0= n-p \geq n-\frac{n(k-1)}{k}=\frac{n}{k}$, $1 \leq k_0\leq k$, and $1 \leq d_0\leq d$. Let $k_0=1$ or $d_0=1$. Using part (a) of the lemma statement, we obtain $h_{AR}(S_{\alpha }^{\prime })\geq n_0 \geq \frac{n}{k}\geq \frac{n}{k}\frac{(k-1)}{k^{d}}=\frac{n(k-1)}{k^{d+1}}$.

Let $k_0\ge 2$ and $d_0 \ge 2$. In this case, $d \ge 2$.
By inductive hypothesis, $h_{AR}(S_{\alpha }^{\prime })\geq \frac{n_0(k_0-1)}{k_0^{d_0}}\ge\frac{n}{k}\frac{(k_0-1)}{k_0^{d}}$. We now show that $\frac{k_0-1}{k_{0}^{d}}\ge \frac{k-1}{k^{d}}$. To this end, for $x \ge 2$, we consider  the function $f(x)=\frac{x-1}{x^{d}}$ and its  derivative $f^{\prime }(x)=\frac{1-d}{x^{d}}+\frac{d}{x^{d+1}}$. It is easy to show that $f^{\prime }(x)<0$ if $x>2$. Using mean value theorem and the inequalities $2 \le k_0 \le k$, we obtain $f(k_0)\ge f(k)$.
Therefore $h_{AR}(S_{\alpha }^{\prime })\geq \frac{n(k-1)}{k^{d+1}}$.
By Lemma \ref{L6}, $h_{AR}(S^{\prime })\geq h_{AR}(S_{\alpha
}^{\prime })$. From the considered relations and (\ref{E2}) it follows that $%
h_{AR}(S)\geq \frac{n(k-1)}{k^{d+1}}$.

Let us assume now that $\left\vert S^{\prime }\right\vert >\frac{n(k-1)}{k}$%
. Denote $m=n-\left\vert S^{\prime }\right\vert $. From (\ref{E3}) it
follows that $m\geq 0$. Let $m=0$. Then, as it is not difficult to note, $%
d(S^{\prime })=1$. Using part (a) of the lemma statement, we obtain  $h_{AR}(S^{\prime })\geq n(S^{\prime }) =\left\vert S^{\prime }\right\vert \geq \frac{n(k-1)}{k}\geq
\frac{n(k-1)}{k^{d+1}}$. Using (\ref{E2}), we obtain $h_{AR}(S)\geq \frac{%
n(k-1)}{k^{d+1}}$.

We now assume that $m>0$. Denote $B=A(S)\setminus
\{a_{i_{1}},\ldots ,a_{i_{p}}\}$. It is clear that $\left\vert B\right\vert
=m$. Let $B=\{a_{l_{1}},\ldots ,a_{l_{m}}\}$ and $l_{1}<\cdots <l_{m}$. Let $%
j\in \{1,\ldots ,m\}$. We now define a set $V_{j}$. If $\left\vert
V_{S^{\prime }}(a_{l_{j}})\right\vert =k$, then $V_{j}=V_{S^{\prime
}}(a_{l_{j}})$. If $\left\vert V_{S^{\prime }}(a_{l_{j}})\right\vert <k$,
then $V_{j}$ is a subset of $\omega $ such that $\left\vert V_{j}\right\vert
=k$ and $V_{S^{\prime }}(a_{l_{j}})\subset V_{j}$. Denote $V=V_{1}\times
\cdots \times V_{m}$. Let $q=d(S^{\prime })-1$. It is clear that $q\leq
m$ and $q\leq d$. It is also clear that, for any decision rule $r\in S^{\prime }$,
there exist at least $k^{m-q}$ tuples $\bar{\delta}=(\delta _{1},\ldots
,\delta _{m})\in V$ such that the system of equations $K(r)\cup
\{a_{l_{1}}=\delta _{1},\ldots ,a_{l_{m}}=\delta _{m}\}$ is consistent. For
each $\bar{\delta}=(\delta _{1},\ldots ,\delta _{m})\in V$, let $N(\bar{%
\delta})$ be the number of decision rules $r\in S^{\prime }$ such that the
system $K(r)\cup \{a_{l_{1}}=\delta _{1},\ldots ,a_{l_{m}}=\delta _{m}\}$ is
consistent. Denote $N=\sum_{\bar{\delta}\in V}N(\bar{\delta})$. It is clear
that $N\geq \left\vert S^{\prime }\right\vert k^{m-q}\geq \frac{k^{m}(k-1)n}{%
k^{q+1}}$. It is also clear that there exists a tuple $\bar{\delta}^{\prime
}\in V$ such that $N(\bar{\delta}^{\prime })\geq \frac{N}{\left\vert
V\right\vert }=\frac{N}{k^{m}}\geq \frac{n(k-1)}{k^{q+1}}\geq \frac{n(k-1)}{%
k^{d+1}}$. Let $\bar{\delta}^{\prime }=(\delta _{1}^{\prime },\ldots ,\delta
_{m}^{\prime })$. We now define a tuple $\bar{\delta}=(\delta _{1},\ldots
,\delta _{m})$. Let $j\in \{1,\ldots ,m\}$. If $\delta _{j}^{\prime }\in
V_{S^{\prime }}(a_{l_{j}})$, then $\delta _{j}=\delta _{j}^{\prime }$. If $%
\delta _{j}^{\prime }\notin V_{S^{\prime }}(a_{l_{j}})$, then $\delta _{j}$
is the minimum number from the set $V_{S^{\prime }}(a_{l_{1}})$. It is not
difficult to show that $N(\bar{\delta})\geq N(\bar{\delta}^{\prime })\geq
\frac{n(k-1)}{k^{d+1}}$. Denote $\alpha =\{a_{l_{1}}=\delta _{1},\ldots
,a_{l_{m}}=\delta _{m}\}$ and consider the system of decision rules $%
S_{\alpha }^{\prime }$. One can show that $k(S_{\alpha }^{\prime })=1$, $%
d(S_{\alpha }^{\prime })=1$, and $n(S_{\alpha }^{\prime })=N(\bar{\delta})\geq \frac{n(k-1)}{%
k^{d+1}}$. Using part (a) of the lemma statement, we obtain $%
h_{AR}(S_{\alpha }^{\prime })\geq \frac{n(k-1)}{k^{d+1}}$. By Lemma \ref{L6}%
, $h_{AR}(S^{\prime })\geq \frac{n(k-1)}{k^{d+1}}$. From this inequality and
(\ref{E2}) it follows that $h_{AR}(S)\geq \frac{n(k-1)}{k^{d+1}}$. Therefore
the inequality (\ref{E1}) holds for $d+1$. Thus, the inequality (\ref{E1})
holds.
\end{proof}

\begin{lemma}
\label{L20} Let $n,d,k\in \omega \setminus \{0\}$, $d\leq n$, $d\geq 2$, and
$k\geq 2$. Then $h_{EAR}(n,d,k)\leq d+\frac{n}{k^{d-1}}$.
\end{lemma}

\begin{proof}
We now describe a system of decision rules $S$ with $A(S)=\{a_{1},\ldots
,a_{n}\}$. Denote $E_{k}=\{0,1,\ldots ,k-1\}$. Let us consider a partition $%
\{a_{d},\ldots ,a_{n}\}=\bigcup_{\bar{\delta}\in E_{k}^{d-1}}B(\bar{\delta})$
such that $\left\vert B(\bar{\delta})\right\vert \leq \left\lceil \frac{n-d+1%
}{k^{d-1}}\right\rceil $ for any $\bar{\delta}\in E_{k}^{d-1}$ (we assume that $B(\bar{\delta}_{1})\cap B(\bar{%
\delta}_{2})=\emptyset $ for any $\bar{\delta}_{1},\bar{\delta}_{2}\in
E_{k}^{d-1}$, $\bar{\delta}_{1}\neq \bar{\delta}_{2}$). It is clear that
some sets in this partition can be empty, but at least one is nonempty. Let $%
\bar{\delta}=(\delta _{1},\ldots ,\delta _{d-1})\in E_{k}^{d-1}$. Describe a
system of decision rules $S(\bar{\delta})$. If $B(\bar{\delta})=\emptyset $,
then $S(\bar{\delta})=\{(a_{1}=\delta _{1})\wedge \cdots \wedge
(a_{d-1}=\delta _{d-1})\rightarrow 0\}$. If $B(\bar{\delta})\neq \emptyset $%
, then $S(\bar{\delta})=\{(a_{1}=\delta _{1})\wedge \cdots \wedge
(a_{d-1}=\delta _{d-1})\wedge (a_{i}=0)\rightarrow 0:a_{i}\in B(\bar{\delta}%
)\}$. Denote $S=\bigcup_{\bar{\delta}\in E_{k}^{d-1}}S(\bar{\delta})$. It is
clear that $n(S)=n$, $d(S)=d$, and $k(S)=k$.

Let us describe a decision tree $\Gamma $. First, we compute the values of
attributes $a_{1},\ldots ,a_{d-1}$. Let $a_{1}=\delta _{1},\ldots
,a_{d-1}=\delta _{d-1}$. If there exists $i\in \{1,\ldots ,d-1\}$ such that $%
\delta _{i}=\ast $, then the output of $\Gamma $ is the empty set of
decision rules. Let $\bar{\delta}=(\delta _{1},\ldots ,\delta _{d-1})\in
E_{k}^{d-1}$. If $B(\bar{\delta})=\emptyset $, then the output of $\Gamma $
is the set $S(\bar{\delta})$. Let $B(\bar{\delta})\neq \emptyset $ and $B(%
\bar{\delta})=\{a_{i_{1}},\ldots ,a_{i_{m}}\}$. We compute the values of
attributes $a_{i_{1}},\ldots ,a_{i_{m}}$. Let $a_{i_{1}}=\sigma _{1},\ldots
,a_{i_{m}}=\sigma _{m}$. Then the output of $\Gamma $ is the set of decision
rules $\{(a_{1}=\delta _{1})\wedge \cdots \wedge (a_{d-1}=\delta
_{d-1})\wedge (a_{i_{j}}=0)\rightarrow 0:j\in \{1,\ldots ,m\},\sigma
_{j}=0\} $. One can show that $\Gamma $ solves the problem $EAR(S)$ and $%
h(\Gamma )\leq d-1+\left\lceil \frac{n-d+1}{k^{d-1}}\right\rceil \leq d+%
\frac{n}{k^{d-1}}$. Therefore $h_{EAR}(n,d,k)\leq d+\frac{n}{k^{d-1}}$.
\end{proof}

\begin{lemma}
\label{L21} Let $n,d,k\in \omega \setminus \{0\}$ and $d\leq n$. Then

(a) If $k=1$ or $d=1$, then $h_{AR}(n,d,k)=h_{EAR}(n,d,k)=n$.

(b) If $k\geq 2$ and $d\geq 2$, then
\begin{equation*}
\max \left\{ d,\frac{n(k-1)}{k^{d}}\right\} \leq h_{AR}(n,d,k)\leq
h_{EAR}(n,d,k)\leq d+\frac{n}{k^{d-1}}.
\end{equation*}
\end{lemma}

\begin{proof}
From Lemma \ref{L14} it follows that $h_{AR}(n,d,k)\leq h_{EAR}(n,d,k)\leq n$%
. Using these inequalities and Lemmas \ref{L19} and \ref{L20}, we obtain the
considered statements.
\end{proof}

\subsection{Bounds on $h_{ESR}^{R}(n,d,k)$ and $h_{EAD}^{R}(n,d,k)$ \label%
{S5.3}}

\begin{lemma}
\label{L22} Let $C\in \{SR,AD\}$, $n,d,k\in \omega \setminus \{0\}$ and $%
d\leq n$. Then $h_{EC}^{R}(n,d,k)\geq \max \left\{ d,\frac{(nk)^{1/d}}{k}%
-d\right\} $.
\end{lemma}

\begin{proof}
Let $S\in \Sigma _{C}$. Using Lemma \ref{L8}, we obtain that $h_{EC}(S)\geq
d(S)$. Since $S$ is an arbitrary $C$-reduced system, the inequality $%
h_{EC}^{R}(n,d,k)\geq d$ holds.

We prove by induction on $d$ that, for any $n,d,k\in \omega \setminus \{0\}$
such that $d\leq n\,$, the following statement holds. Let $S^{\prime }\in
\Sigma _{C}$, $S\subseteq S^{\prime }$, $n(S)=n$, $k(S)=k$, and $d(S)=d$.
Then
\[
h_{EC}(S^{\prime })\geq \frac{(nk)^{1/d}}{k}-d.
\]%

Let $S^{\prime }\in \Sigma _{C}$, $S\subseteq S^{\prime }$, $n(S)=n$, $k(S)=k
$, and $d(S)=1$. It is clear that $\beta (S^{\prime })\geq \beta (S)=n$. Let
us show that $I_{C}(S^{\prime })=S^{\prime }$ (corresponding definitions are
before Lemma \ref{L7}). If $C=SR$, then $S^{\prime }$ does not contain both
a rule of the length $0$ and a rule of the length greater than $0$. If $C=AD$%
, then $S^{\prime }$ does not contain both a rule of the length $0$ and a
rule of the length greater than $0$ with the same right-hand sides. Using
Remarks \ref{R1} and \ref{R2}, we obtain $I_{C}(S^{\prime })=S^{\prime }$.
By Lemma \ref{L7}, $h_{EC}(S^{\prime })\geq n$. Thus, for $d=1$, the
considered statement holds. 

Let us assume that, for some $d\geq 1$, this
statement holds for each $d^{\prime }$, $1\leq d^{\prime }\leq d$. We prove
that it holds for $d+1$. Let $S^{\prime }\in \Sigma _{C}$, $S\subseteq
S^{\prime }$, $n(S)=n$, $k(S)=k$, and $d(S)=d+1$. We now show that
\begin{equation}
h_{EC}(S^{\prime })\geq \frac{(nk)^{\frac{1}{d+1}}}{k}-d-1.  \label{E5}
\end{equation}%

By proved above, $h_{EC}(S^{\prime })\geq d(S^{\prime })\geq d(S)\geq 2$.
Therefore if $2\geq \frac{(nk)^{\frac{1}{d+1}}}{k}-d-1$, then the inequality
(\ref{E5}) holds. Let us assume that $\frac{(nk)^{\frac{1}{d+1}}}{k}-d-1>2$.
Then%
\begin{equation}
\frac{(nk)^{\frac{1}{d+1}}}{k}>2.  \label{E6}
\end{equation}

Denote $m=\beta (S)$. It is clear that $\beta (S^{\prime })\geq \beta (S)$.
Using Remarks \ref{R1} and \ref{R2} and the fact that $S^{\prime }\in \Sigma
_{C}$, we obtain $I_{C}(S^{\prime })=S^{\prime }$. From Lemma \ref{L7} it
follows that $h_{EC}(S^{\prime })\geq m$. Let $m\geq $ $\frac{(nk)^{\frac{1}{%
d+1}}}{k}-d-1$. Then the inequality (\ref{E5}) holds. Let us assume now that
$\frac{(nk)^{\frac{1}{d+1}}}{k}-d-1>m$. Then
\begin{equation}
\frac{(nk)^{\frac{1}{d+1}}}{k}>m.  \label{E7}
\end{equation}%

Let $B=\{a_{i_{1}},\ldots ,a_{i_{m}}\}$ be a node cover of the hypergraph $%
G(S)$ with the minimum cardinality. For arbitrary $j\in \{1,\ldots ,m\}$ and
$\delta \in V_{S}(a_{i_{j}})$, we denote by $S(a_{i_{j}}=\delta )$ the set
of decision rules from $S$, which have the equation $a_{i_{j}}=\delta $ in
the left-hand side. Since $B$ is a node cover of the hypergraph $G(S)$, $%
A(S)=\bigcup_{\substack{ j\in \{1,\ldots ,m\} \\ \delta \in V_{S}(a_{i_{j}})
}}A(S(a_{i_{j}}=\delta ))$. Therefore there exist $j\in \{1,\ldots ,m\}$ and
$\delta \in V_{S}(a_{i_{j}})$ such that $\left\vert A(S(a_{i_{j}}=\delta
))\right\vert \geq \frac{n}{mk}$. Using (\ref{E7}), we obtain%
\begin{equation}
\left\vert A(S(a_{i_{j}}=\delta ))\right\vert >\frac{n}{(nk)^{\frac{1}{d+1}}}%
.  \label{E8}
\end{equation}%

Denote $\alpha =\{a_{i_{j}}=\delta \}$. Let us consider the system of
decision rules $S^{\prime \prime }=R_{C}(S_{\alpha }^{\prime })$. Denote by $%
S_{0}$ the system of decision rules obtained from the system $%
S(a_{i_{j}}=\delta )$ by removing the equation $a_{i_{j}}=\delta $ from the
left-hand sides of the rules included in $S(a_{i_{j}}=\delta )$. Denote $%
k_{0}=k(S_{0})$, $d_{0}=d(S_{0})$, and $n_{0}=n(S_{0})$. One can show that $%
k_{0}\leq k$, $d_{0}\leq d$, $n_{0}=\left\vert A(S(a_{i_{j}}=\delta
))\right\vert -1$, and $S_{0}\subseteq S^{\prime \prime }$. Using (\ref{E8}%
), we obtain that $n_{0}>\frac{n}{(nk)^{\frac{1}{d+1}}}-1$. Denote $q=\frac{n%
}{(nk)^{\frac{1}{d+1}}}$. From (\ref{E6}) it follows that $n>2^{d+1}k^{d}$.
As a result, we have $q=\frac{n^{\frac{d}{d+1}}}{k^{\frac{1}{d+1}}}>\frac{%
(2^{d+1}k^{d})^{\frac{d}{d+1}}}{k^{\frac{1}{d+1}}}=2^{d}k^{\frac{d^{2}-1}{d+1%
}}=2^{d}k^{d-1}\geq 2^{d}\geq 2$. Using the inductive hypothesis, we obtain
\[
h_{EC}(S^{\prime \prime })\geq \frac{(n_{0}k_{0})^{1/d_{0}}}{k_{0}}%
-d_{0}\geq \frac{((q-1)k)^{1/d}}{k}-d=\frac{((q-1)k)^{1/d}}{k}+1-d-1.
\]%
By Lemma \ref{L9}, $h_{EC}(S^{\prime \prime })=h_{EC}(S_{\alpha }^{\prime })$%
. From  Lemma \ref{L6} it follows that $h_{EC}(S^{\prime })\geq
h_{EC}(S_{\alpha }^{\prime })=h_{EC}(S^{\prime \prime })$. We now prove
that $\frac{((q-1)k)^{\frac{1}{d}}}{k}+1\geq \frac{(qk)^{\frac{1}{d}}}{k}$.
To this end, we should show that $(q-1)^{\frac{1}{d}}+k^{1-\frac{1}{d}}-q^{%
\frac{1}{d}}\geq 0$. It is clear that $k^{1-\frac{1}{d}}\geq 1$. We now
prove that $(q-1)^{\frac{1}{d}}+1-q^{\frac{1}{d}}\geq 0$. For this, for $%
x\geq 2$, we consider the function $f(x)=$ $(x-1)^{\frac{1}{d}}+1-x^{\frac{1%
}{d}}$ and its derivative $f^{\prime }(x)=\frac{1}{d(x-1)^{\frac{d-1}{d}}}-%
\frac{1}{dx^{\frac{d-1}{d}}}$. It is easy to show that $f^{\prime }(x)>0$
for any $x\geq 2$. Using mean value theorem, we obtain that $f(x)\geq f(2)$
for any $x\geq 2$. It is clear that $f(2)=2-2^{\frac{1}{d}}\geq 0$.
Therefore $h_{EC}(S^{\prime })\geq \frac{(qk)^{\frac{1}{d}}}{k}-d-1$. It is
easy to show that $\frac{(qk)^{1/d}}{k}=\frac{(nk)^{\frac{1}{d+1}}}{k}$.
Hence $h_{EC}(S^{\prime })\geq \frac{(nk)^{\frac{1}{d+1}}}{k}-d-1$. Thus,
the inequality (\ref{E5}) is proved. Therefore the considered statement holds.
From this statement it follows immediately that, for any $n,d,k\in \omega
\setminus \{0\}$ such that $d\leq n$, the inequality $h_{EC}^{R}(n,d,k)\geq
\frac{(nk)^{1/d}}{k}-d$ holds.
\end{proof}

\begin{lemma}
\label{L23} Let $n,d,k\in \omega \setminus \{0\}$ and $d\leq n$. Then $%
h_{ESR}^{R}(n,d,k)\leq 2d\left\lceil \frac{(nk)^{1/d}}{k}\right\rceil $ and $%
h_{EAD}^{R}(n,d,k)\leq 2d\left\lceil \frac{(nk)^{1/d}}{k}\right\rceil $.
\end{lemma}

\begin{proof}
We prove by induction on $d$ that, for any $n,d,k\in \omega \setminus \{0\}$
such that $d\leq n$, there exists an $SR$-reduced system of decision rules $%
S $ satisfying the following conditions: the right-hand sides of all rules
from $S$ are equal to $0$, $n(S)=n$, $d(S)=d$, $k(S)=k$, and $h_{ESR}(S)\leq
2d\left\lceil \frac{(nk)^{1/d}}{k}\right\rceil $.

Let $d=1$. Consider the system $S=\{(a_{1}=0)\rightarrow
0,(a_{1}=1)\rightarrow 0,\ldots ,(a_{1}=k-1)\rightarrow
0,(a_{2}=0)\rightarrow 0,\ldots ,(a_{n}=0)\rightarrow 0\}$. It is clear that
$S\in \Sigma _{SR}$, $n(S)=n$, $d(S)=1$, $k(S)=k$, and the right-hand sides of
all rules from $S$ are equal to $0$. Using Lemma \ref{L14}, we obtain that $%
h_{ESR}(S)\leq n$. Therefore, for $d=1$, the considered statement holds. Let
us assume that this statement holds for some $d\geq 1$. We will show that it
also holds for $d+1$.

Denote $t=\left\lceil \frac{(nk)^{\frac{1}{d+1}}}{k}\right\rceil $. It is
clear that $t\geq 1$. Let $2(d+1)t\geq n$. Let us consider a system of
decision rules $S=\{(a_{1}=0)\wedge \cdots \wedge (a_{d+1}=0)\rightarrow
0,(a_{1}=1)\rightarrow 0,\ldots ,(a_{1}=k-1)\rightarrow
0,(a_{d+2}=0)\rightarrow 0,\ldots ,(a_{n}=0)\rightarrow 0\}$. It is clear
that $n(S)=n$, $d(S)=d+1$, $k(S)=k$, $S\in \Sigma _{SR}$, and the right-hand
sides of all rules from $S$ are equal to $0$. Using Lemma \ref{L14}, we
obtain that $h_{ESR}(S)\leq n\leq 2(d+1)t$. Therefore in this case the
considered statement holds.

Let us assume now that $2(d+1)t<n$. Then $%
n-2t>2dt\geq 2d.$ Using these inequalities, we obtain that there exist
numbers $n_{0},n_{1},\ldots ,n_{2tk}\in \omega $ satisfying the following
conditions: $\sum_{i=0}^{2tk}n_{i}=n$, $n_{0}=2t$, $n_{1}=d$, if $%
\left\lceil \frac{n-2t}{2tk}\right\rceil <d$, and $d$ $\leq n_{1}\leq
\left\lceil \frac{n-2t}{2tk}\right\rceil $ otherwise, and $n_{i}$ $\leq
\left\lceil \frac{n-2t}{2tk}\right\rceil $ for $i=2,\ldots ,2tk$. It is
clear that there exist sets of attributes $A_{0}^{\prime },A_{1}^{\prime
},\ldots ,A_{2tk}^{\prime }$ such that $A_{i}^{\prime }\cap A_{j}^{\prime
}=\emptyset $ for any $i,j\in \{0,1,\ldots ,2tk\}$, $i\neq j$, $\left\vert
A_{i}^{\prime }\right\vert =n_{i}$ for $i=0,1,\ldots ,2tk$, $%
\bigcup_{i=0}^{2tk}A_{i}^{\prime }=\{a_{1},\ldots ,a_{n}\}$, and $%
A_{0}^{\prime }=\{a_{1},\ldots ,a_{2t}\}$. We now define sets of attributes $%
A_{0},A_{1},\ldots ,A_{2tk}$. Let $i\in \{0,1,\ldots ,2tk\}$. If $%
A_{i}^{\prime }\neq \emptyset $, then $A_{i}=A_{i}^{\prime }$. If $%
A_{i}^{\prime }=\emptyset $, then $A_{i}=\{a_{n}\}$. It is clear that $%
A_{0}=\{a_{1},\ldots ,a_{2t}\}$, $\left\vert A_{1}\right\vert \geq d$, $%
A_{0}\cap A_{i}=\emptyset $ and $A_{i}\neq \emptyset $ for $i=1,\ldots ,2tk$%
, $\bigcup_{i=0}^{2tk}A_{i}=\{a_{1},\ldots ,a_{n}\}$, and, for $i=2,\ldots ,2tk$,
\begin{equation}
\left\vert A_{i}\right\vert \leq \left\lceil \frac{n-2t}{2tk}\right\rceil .
\label{E10}
\end{equation}%

For each $i\in \{1,\ldots ,2tk\}$, define a decision rule system $S_{i}$
such that $A(S_{i})=A_{i}$. Let $\left\vert A_{i}\right\vert \leq d$ and $%
A_{i}=\{a_{j_{1}},\ldots ,a_{j_{m}}\}$. Then $S_{i}=\{(a_{j_{1}}=0)\wedge
\cdots \wedge (a_{j_{m}}=0)\rightarrow 0,(a_{j_{1}}=1)\rightarrow 0,\ldots
,(a_{j_{1}}=k-1)\rightarrow 0\}$. It is clear that $n(S_{i})=\left\vert
A_{i}\right\vert $, $d(S_{i})=\left\vert A_{i}\right\vert $, $k(S_{i})=k$,
the right-hand sides of decision rules from $S_{i}$ are equal to $0$, and $%
S_{i}\in \Sigma _{SR}$. Using Lemma \ref{L14}, we obtain that $%
h_{ESR}(S_{i})\leq m\leq d$. Therefore %
\begin{equation}
h_{ESR}(S_{i})\leq 2d\left\lceil \frac{(nk)^{\frac{1}{d+1}}}{k}\right\rceil .
\label{E11}
\end{equation}%

Let $\left\vert A_{i}\right\vert >d$. Then $\left\vert A_{i}\right\vert \geq
2$. Using inductive hypothesis, we obtain that there exists a decision rule
system $S_{i}$ satisfying the following conditions: $n(S_{i})=\left\vert
A_{i}\right\vert $, $d(S_{i})=d$, $k(S_{i})=k$, $S_{i}\in \Sigma _{SR}$, the
right-hand sides of decision rules from $S_{i}$ are equal to $0$, and
\begin{equation}
h_{ESR}(S_{i})\leq 2d\left\lceil \frac{(\left\vert A_{i}\right\vert k)^{1/d}%
}{k}\right\rceil .  \label{E12}
\end{equation}%
We now show that the system $S_{i}$ satisfies the inequality (\ref{E11}).
Using the inequality (\ref{E10}), we obtain that
\begin{equation*}
2\leq \left\vert A_{i}\right\vert \leq \left\lceil \frac{n-2t}{2tk}%
\right\rceil \leq \frac{n}{2tk}+1\leq \frac{n}{tk}\leq \frac{nk}{k(nk)^{%
\frac{1}{d+1}}}=\frac{n}{(nk)^{\frac{1}{d+1}}}.
\end{equation*}%
From these relations and the inequality (\ref{E12}) it follows that
\begin{equation*}
h_{ESR}(S_{i})\leq 2d\left\lceil \frac{\left( \frac{nk}{(nk)^{\frac{1}{d+1}}}%
\right) ^{1/d}}{k}\right\rceil =2d\left\lceil \frac{(nk)^{\frac{1}{d+1}}}{k}%
\right\rceil .
\end{equation*}%
Thus, the inequality (\ref{E11}) holds.

Let $j\in \{1,\ldots ,2t\}$ and $\sigma \in \{0,\ldots ,k-1\}$. Describe a
decision rule system $S_{j\sigma }$. Let $i=k(j-1)+\sigma +1$. We add to the
left-hand side of each rule from $S_{i}$ the equation $a_{j}=\sigma $. We
denote $S_{j\sigma }$ the obtained decision rule system. Denote $%
S_{0}=\{(a_{i}=\delta )\wedge (a_{j}=\sigma )\rightarrow 0:1\leq i<j\leq
2t;\delta ,\sigma \in \{0,\ldots ,k-1\}\}$. We now consider the decision
rule system%
\begin{equation*}
S=S_{0}\cup \bigcup_{\substack{ j\in \{1,\ldots ,2t\}  \\ \sigma \in
\{0,\ldots ,k-1\}}}S_{j\sigma .}
\end{equation*}%
One can show that $n(S)=n$, $d(S)=d+1$, $k(S)=k$, $S\in \Sigma _{SR}$, and
the right-hand sides of all rules from $S$ are equal to $0$.

Let us describe the operation of
a decision tree $\Gamma $ solving the problem $ESR(S)$. First, we compute
the values of the attributes $a_{1},\ldots ,a_{2t}$. If $a_{1}=\cdots
=a_{2t}=\ast $, then no one rule from $S$ is realizable for the considered
tuple of attribute values. Let there exist $i,j\in \{1,\ldots ,2t\}$ such
that $a_{i}=\delta $ and $a_{j}=\sigma $, where $i\neq j$ and $\sigma
,\delta \in \{0,\ldots ,k-1\}$. Then, for the considered tuple of attribute
values, the rule $(a_{i}=\delta )\wedge (a_{j}=\sigma )\rightarrow 0$ from $%
S $ is realizable. Let $a_{1}=\cdots =a_{j-1}=a_{j+1}=\cdots =a_{2t}=\ast $
and $a_{j}=\sigma $, where $\sigma \in \{0,\ldots ,k-1\}$. In this case, we
obtain that, for the considered tuple of attribute values, only rules from $%
S_{j\sigma }$ can be realizable. Taking into account that $a_{j}=\sigma $,
we obtain that the problem $ESR(S)$ solving is reduced to the problem $%
ESR(S_{i})$ solving, where $i=k(j-1)+\sigma +1$. We solve the problem $%
ESR(S_{i})$ using a decision tree with the minimum depth. By (\ref{E11}),%
\begin{equation*}
h(\Gamma )\leq 2t+2dt=2(d+1)\left\lceil \frac{(nk)^{\frac{1}{d+1}}}{k}%
\right\rceil .
\end{equation*}%
Hence the considered statement is fully proven.

Let $n,d,k\in \omega  \setminus \{0\}$ and $d\leq n$. Then there exists a decision rule
system $S$ satisfying the following conditions: $n(S)=n$, $d(S)=d$, $k(S)=k$%
, $S\in \Sigma _{SR}$, the right-hand sides of all decision rules from $S$
are equal to $0$, and $h_{ESR}(S)\leq 2d\left\lceil \frac{(nk)^{1/d}}{k}%
\right\rceil $. It is clear that $S\in \Sigma _{AD}$ and $%
h_{EAD}(S)=h_{ESR}(S)$. Therefore the following two inequalities hold:%
\begin{equation*}
h_{ESR}^{R}(n,d,k)\leq 2d\left\lceil \frac{(nk)^{1/d}}{k}\right\rceil
, h_{EAD}^{R}(n,d,k)\leq 2d\left\lceil \frac{(nk)^{1/d}}{k}\right\rceil .
\end{equation*}

\end{proof}

\subsection{Theorem on Lower Bounds \label{S5.4}}

\begin{theorem}
\label{T2} Let $n,d,k\in \omega \setminus \{0\}$ and $d\leq n$. Then

(a) $h_{SR}(n,d,k)=h_{AD}(n,d,k)=h_{ESR}(n,d,k)=h_{EAD}(n,d,k)=0$.

(b) If $k=1$ or $d=1$, then $h_{AR}(n,d,k)=h_{EAR}(n,d,k)=n$. If $k\geq 2$
and $d\geq 2$, then
\begin{equation*}
\max \left\{ d,\frac{n(k-1)}{k^{d}}\right\} \leq h_{AR}(n,d,k)\leq
h_{EAR}(n,d,k)\leq d+\frac{n}{k^{d-1}}.
\end{equation*}

(c) $h_{SR}^{R}(n,d,k)=h_{AD}^{R}(n,d,k)=\left\{
\begin{array}{ll}
n, & \text{if }d=n, \\
1, & \text{if }d\neq n.%
\end{array}%
\right. $

(d) $\max \left\{ d,\frac{(nk)^{1/d}}{k}-d\right\} \leq h_{ESR}^{R}(n,d,k)\leq
2d\left\lceil \frac{(nk)^{1/d}}{k}\right\rceil .$

(e) $\max \left\{ d,\frac{(nk)^{1/d}}{k}-d\right\} \leq h_{EAD}^{R}(n,d,k)\leq
2d\left\lceil \frac{(nk)^{1/d}}{k}\right\rceil .$
\end{theorem}

\begin{proof}
The statements of the theorem follow from Lemmas \ref{L16}, \ref{L17}, \ref%
{L21}, \ref{L22}, and \ref{L23}.
\end{proof}

\begin{remark}
Note that the results for $h_{AR}(n,d,k)$ mentioned in the theorem were
obtained in \cite{Moshkov98}. The results for $h_{EAR}(n,d,k)$ were
published in \cite{Moshkov01} without proof.
\end{remark}

\section{Conclusion\label{S6}}
In this paper, we obtained unimprovable upper and lower bounds of the minimum depth of the decision trees, which solve the problems All Rules (AR), All Decisions (AD), Some Rules (SR) and their extended versions (EAR, EAD, and ESR).

In the future work, we will consider the problems of
constructing decision trees and acyclic decision graphs representing trees for given decision rule systems, and  discuss the possibility of constructing not the entire decision tree, but the computation path in this tree for the given input.

\subsection*{Acknowledgements}

Research reported in this publication was supported by King Abdullah
University of Science and Technology (KAUST).

\bibliographystyle{spmpsci}
\bibliography{1C}

\begin{thebibliography}{10}
\providecommand{\url}[1]{{#1}}
\providecommand{\urlprefix}{URL }
\expandafter\ifx\csname urlstyle\endcsname\relax
  \providecommand{\doi}[1]{DOI~\discretionary{}{}{}#1}\else
  \providecommand{\doi}{DOI~\discretionary{}{}{}\begingroup
  \urlstyle{rm}\Url}\fi

\bibitem{AbdelhalimTN16}
Abdelhalim, A., Traor{\'{e}}, I., Nakkabi, Y.: Creating decision trees from
  rules using {RBDT-1}.
\newblock Comput. Intell. \textbf{32}(2), 216--239 (2016)

\bibitem{AbdelhalimTS09}
Abdelhalim, A., Traor{\'{e}}, I., Sayed, B.: {RBDT-1:} {A} new rule-based
  decision tree generation technique.
\newblock In: G.~Governatori, J.~Hall, A.~Paschke (eds.) Rule Interchange and
  Applications, International Symposium, RuleML 2009, Las Vegas, Nevada, USA,
  November 5-7, 2009. Proceedings, \emph{Lecture Notes in Computer Science},
  vol. 5858, pp. 108--121. Springer (2009)

\bibitem{AbouEishaACHM19}
AbouEisha, H., Amin, T., Chikalov, I., Hussain, S., Moshkov, M.: Extensions of
  Dynamic Programming for Combinatorial Optimization and Data Mining,
  \emph{Intelligent Systems Reference Library}, vol. 146.
\newblock Springer (2019)

\bibitem{AlsolamiACM20}
Alsolami, F., Azad, M., Chikalov, I., Moshkov, M.: Decision and Inhibitory
  Trees and Rules for Decision Tables with Many-valued Decisions,
  \emph{Intelligent Systems Reference Library}, vol. 156.
\newblock Springer (2020)

\bibitem{BlumI87}
Blum, M., Impagliazzo, R.: Generic oracles and oracle classes (extended
  abstract).
\newblock In: 28th Annual Symposium on Foundations of Computer Science, Los
  Angeles, California, USA, 27-29 October 1987, pp. 118--126. {IEEE} Computer
  Society (1987)

\bibitem{BorosHIK97}
Boros, E., Hammer, P.L., Ibaraki, T., Kogan, A.: Logical analysis of numerical
  data.
\newblock Math. Program. \textbf{79}, 163--190 (1997)

\bibitem{BorosHIKMM00}
Boros, E., Hammer, P.L., Ibaraki, T., Kogan, A., Mayoraz, E., Muchnik, I.B.: An
  implementation of logical analysis of data.
\newblock {IEEE} Trans. Knowl. Data Eng. \textbf{12}(2), 292--306 (2000)

\bibitem{BreimanFOS84}
Breiman, L., Friedman, J.H., Olshen, R.A., Stone, C.J.: {Classification and
  Regression Trees}.
\newblock Wadsworth and Brooks (1984)

\bibitem{BuhrmanW02}
Buhrman, H., de~Wolf, R.: Complexity measures and decision tree complexity: a
  survey.
\newblock Theor. Comput. Sci. \textbf{288}(1), 21--43 (2002)

\bibitem{CaoSJ20}
Cao, H.E.C., Sarlin, R., Jung, A.: Learning explainable decision rules via
  maximum satisfiability.
\newblock {IEEE} Access \textbf{8}, 218180--218185 (2020)

\bibitem{ChikalovLLMNSZ13}
Chikalov, I., Lozin, V.V., Lozina, I., Moshkov, M., Nguyen, H.S., Skowron, A.,
  Zielosko, B.: Three Approaches to Data Analysis - Test Theory, Rough Sets and
  Logical Analysis of Data, \emph{Intelligent Systems Reference Library},
  vol.~41.
\newblock Springer (2013)

\bibitem{FurnkranzGL12}
F{\"{u}}rnkranz, J., Gamberger, D., Lavrac, N.: Foundations of Rule Learning.
\newblock Cognitive Technologies. Springer (2012)

\bibitem{GilmoreEH21}
Gilmore, E., Estivill{-}Castro, V., Hexel, R.: More interpretable decision
  trees.
\newblock In: H.~Sanjurjo{-}Gonz{\'{a}}lez, I.~Pastor{-}L{\'{o}}pez, P.G.
  Bringas, H.~Quinti{\'{a}}n, E.~Corchado (eds.) Hybrid Artificial Intelligent
  Systems - 16th International Conference, {HAIS} 2021, Bilbao, Spain,
  September 22-24, 2021, Proceedings, \emph{Lecture Notes in Computer Science},
  vol. 12886, pp. 280--292. Springer (2021)

\bibitem{HartmanisH87}
Hartmanis, J., Hemachandra, L.A.: One-way functions, robustness, and the
  non-isomorphism of {NP}-complete sets.
\newblock In: Proceedings of the Second Annual Conference on Structure in
  Complexity Theory, Cornell University, Ithaca, New York, USA, June 16-19,
  1987. {IEEE} Computer Society (1987)

\bibitem{ImamM93a}
Imam, I.F., Michalski, R.S.: Learning decision trees from decision rules: {A}
  method and initial results from a comparative study.
\newblock J. Intell. Inf. Syst. \textbf{2}(3), 279--304 (1993)

\bibitem{ImamM93}
Imam, I.F., Michalski, R.S.: Should decision trees be learned from examples of
  from decision rules?
\newblock In: H.J. Komorowski, Z.W. Ras (eds.) Methodologies for Intelligent
  Systems, 7th International Symposium, {ISMIS} '93, Trondheim, Norway, June
  15-18, 1993, Proceedings, \emph{Lecture Notes in Computer Science}, vol. 689,
  pp. 395--404. Springer (1993)

\bibitem{ImamM96}
Imam, I.F., Michalski, R.S.: Learning for decision making: the {FRD} approach
  and a comparative study.
\newblock In: Z.W. Ras, M.~Michalewicz (eds.) Foundations of Intelligent
  Systems, 9th International Symposium, {ISMIS} '96, Zakopane, Poland, June
  9-13, 1996, Proceedings, \emph{Lecture Notes in Computer Science}, vol. 1079,
  pp. 428--437. Springer (1996)

\bibitem{KaufmanMPW06}
Kaufman, K.A., Michalski, R.S., Pietrzykowski, J., Wojtusiak, J.: An integrated
  multi-task inductive database {VINLEN:} initial implementation and early
  results.
\newblock In: S.~Dzeroski, J.~Struyf (eds.) Knowledge Discovery in Inductive
  Databases, 5th International Workshop, {KDID} 2006, Berlin, Germany,
  September 18, 2006, Revised Selected and Invited Papers, \emph{Lecture Notes
  in Computer Science}, vol. 4747, pp. 116--133. Springer (2006)

\bibitem{MichalskiI94}
Michalski, R.S., Imam, I.F.: Learning problem-oriented decision structures from
  decision rules: The {AQDT-2} system.
\newblock In: Z.W. Ras, M.~Zemankova (eds.) Methodologies for Intelligent
  Systems, 8th International Symposium, {ISMIS} '94, Charlotte, North Carolina,
  USA, October 16-19, 1994, Proceedings, \emph{Lecture Notes in Computer
  Science}, vol. 869, pp. 416--426. Springer (1994)

\bibitem{MichalskiI97}
Michalski, R.S., Imam, I.F.: On learning decision structures.
\newblock Fundam. Informaticae \textbf{31}(1), 49--64 (1997)

\bibitem{Molnar22}
Molnar, C.: Interpretable Machine Learning. A Guide for Making Black Box Models
  Explainable, 2 edn. (2022).
\newblock \urlprefix\url{christophm.github.io/interpretable-ml-book/}

\bibitem{Moshkov95}
Moshkov, M.: About the depth of decision trees computing {B}oolean functions.
\newblock Fundam. Informaticae \textbf{22}(3), 203--215 (1995)

\bibitem{Moshkov96}
Moshkov, M.: Comparative analysis of deterministic and nondeterministic
  decision tree complexity. {G}lobal approach.
\newblock Fundam. Informaticae \textbf{25}(2), 201--214 (1996)

\bibitem{Moshkov98}
Moshkov, M.: Some relationships between decision trees and decision rule
  systems.
\newblock In: L.~Polkowski, A.~Skowron (eds.) Rough Sets and Current Trends in
  Computing, First International Conference, RSCTC'98, Warsaw, Poland, June
  22-26, 1998, Proceedings, \emph{Lecture Notes in Computer Science}, vol.
  1424, pp. 499--505. Springer (1998)

\bibitem{Moshkov00}
Moshkov, M.: Deterministic and nondeterministic decision trees for rough
  computing.
\newblock Fundam. Informaticae \textbf{41}(3), 301--311 (2000)

\bibitem{Moshkov01}
Moshkov, M.: On transformation of decision rule systems into decision trees (in
  {R}ussian).
\newblock In: Proceedings of the Seventh International Workshop Discrete
  Mathematics and its Applications, Moscow, Russia, January 29 -- February 2,
  2001, Part 1, pp. 21--26. Center for Applied Investigations of Faculty of
  Mathematics and Mechanics, Moscow State University (2001)

\bibitem{Moshkov03}
Moshkov, M.: Classification of infinite information systems depending on
  complexity of decision trees and decision rule systems.
\newblock Fundam. Informaticae \textbf{54}(4), 345--368 (2003)

\bibitem{Moshkov05a}
Moshkov, M.: Comparative analysis of deterministic and nondeterministic
  decision tree complexity. {L}ocal approach.
\newblock In: J.F. Peters, A.~Skowron (eds.) Trans. Rough Sets {IV},
  \emph{Lecture Notes in Computer Science}, vol. 3700, pp. 125--143. Springer
  (2005)

\bibitem{Moshkov05}
Moshkov, M.: Time complexity of decision trees.
\newblock In: J.F. Peters, A.~Skowron (eds.) Trans. Rough Sets III,
  \emph{Lecture Notes in Computer Science}, vol. 3400, pp. 244--459. Springer
  (2005)

\bibitem{Moshkov20}
Moshkov, M.: Comparative Analysis of Deterministic and Nondeterministic
  Decision Trees, \emph{Intelligent Systems Reference Library}, vol. 179.
\newblock Springer (2020)

\bibitem{MPZ08}
Moshkov, M., Piliszczuk, M., Zielosko, B.: Partial Covers, Reducts and Decision
  Rules in Rough Sets - Theory and Applications, \emph{Studies in Computational
  Intelligence}, vol. 145.
\newblock Springer (2008)

\bibitem{MoshkovZ11}
Moshkov, M., Zielosko, B.: Combinatorial Machine Learning - {A} Rough Set
  Approach, \emph{Studies in Computational Intelligence}, vol. 360.
\newblock Springer (2011)

\bibitem{Pawlak91}
Pawlak, Z.: Rough Sets - Theoretical Aspects of Reasoning about Data,
  \emph{Theory and Decision Library: Series {D}}, vol.~9.
\newblock Kluwer (1991)

\bibitem{PawlakS07}
Pawlak, Z., Skowron, A.: Rudiments of rough sets.
\newblock Inf. Sci. \textbf{177}(1), 3--27 (2007)

\bibitem{Quinlan87}
Quinlan, J.R.: Generating production rules from decision trees.
\newblock In: J.P. McDermott (ed.) Proceedings of the 10th International Joint
  Conference on Artificial Intelligence. Milan, Italy, August 23-28, 1987, pp.
  304--307. Morgan Kaufmann (1987)

\bibitem{Quinlan93}
Quinlan, J.R.: {C4.5:} Programs for Machine Learning.
\newblock Morgan Kaufmann (1993)

\bibitem{Quinlan99}
Quinlan, J.R.: Simplifying decision trees.
\newblock Int. J. Hum. Comput. Stud. \textbf{51}(2), 497--510 (1999)

\bibitem{RokachM07}
Rokach, L., Maimon, O.: Data Mining with Decision Trees - Theory and
  Applications, \emph{Series in Machine Perception and Artificial
  Intelligence}, vol.~69.
\newblock World Scientific (2007)

\bibitem{SilvaGKJS20}
Silva, A., Gombolay, M.C., Killian, T.W., Jimenez, I.D.J., Son, S.:
  Optimization methods for interpretable differentiable decision trees applied
  to reinforcement learning.
\newblock In: S.~Chiappa, R.~Calandra (eds.) The 23rd International Conference
  on Artificial Intelligence and Statistics, {AISTATS} 2020, 26-28 August 2020,
  Online [Palermo, Sicily, Italy], \emph{Proceedings of Machine Learning
  Research}, vol. 108, pp. 1855--1865. {PMLR} (2020)

\bibitem{SzydloSM05}
Szydlo, T., Sniezynski, B., Michalski, R.S.: A rules-to-trees conversion in the
  inductive database system {VINLEN}.
\newblock In: M.A. Klopotek, S.T. Wierzchon, K.~Trojanowski (eds.) Intelligent
  Information Processing and Web Mining, Proceedings of the International
  {IIS:} IIPWM'05 Conference held in Gdansk, Poland, June 13-16, 2005,
  \emph{Advances in Soft Computing}, vol.~31, pp. 496--500. Springer (2005)

\bibitem{Tardos89}
Tardos, G.: Query complexity, or why is it difficult to separate ${NP}^{A}\cap
  co{NP}^{A}$ from ${P}^{A}$ by random oracles ${A}$?
\newblock Comb. \textbf{9}(4), 385--392 (1989)

\end{thebibliography}

\end{document}